\documentclass[fleqn,usenatbib]{mnras}

\usepackage{newtxtext,newtxmath}

\usepackage[T1]{fontenc}
\usepackage{ae,aecompl}

\usepackage{graphicx}	
\usepackage{amsmath}	
\usepackage{pdflscape}	
\usepackage{threeparttable}
\usepackage{ulem} 

\newcommand{\tess}{\textit{TESS}}
\newcommand{\target}{HD\,152843}
\newcommand{\planetb}{HD\,152843~b}
\newcommand{\planetc}{HD\,152843~c}
\newcommand{\vespa}{\texttt{VESPA}}

\newcommand\vsini{$v$\,sin\,$i_\star$}

\newcommand{\kms}{\,km\,s$^{-1}$} 
\newcommand{\ms}{\,m\,s$^{-1}$} 
\newcommand{\Msun}{\,$M_{\odot}$\,}	
\newcommand{\Rsun}{\,$R_{\odot}$\,}
\newcommand{\REarth}{\,$R_{\oplus}$\,}

\newcommand{\RMbLC}[1][${\rm m\,s^{-1}}$]   {$3.71 _{ - 0.74 } ^ { + 0.89 }$~#1} 
\newcommand{\RMcLC}[1][${\rm m\,s^{-1}}$]   {$9.56 _{ - 2.7 } ^ { + 2.65 }$~#1} 

\newcommand{\Tzerob}[1][days]   {$1994.2831 _{ - 0.0029 } ^ { + 0.0024 }$~#1} 
\newcommand{\Pb}[1][days]   {$11.6264 _{ - 0.0025 } ^ { + 0.0022 }$~#1} 
\newcommand{\esinb}[1][ ]   {$-0.11 _{ - 0.28 } ^ { + 0.19 }$~#1} 
\newcommand{\ecosb}[1][ ]   {$-0.07 _{ - 0.38 } ^ { + 0.37 }$~#1} 
\newcommand{\bb}[1][ ]   {$0.32 _{ - 0.20 } ^ { + 0.27 }$~#1} 
 
\newcommand{\rrb}[1][ ]   {$0.02201 _{ - 0.00073 } ^ { + 0.00081 }$~#1} 
\newcommand{\kb}[1][${\rm m\,s^{-1}}$]   {$3.09 _{ - 1.66 } ^ { + 1.76 }$~#1} 
\newcommand{\mpb}[1][$M_{\oplus}$]   {$11.56 _{ - 6.14 } ^ { + 6.58 }$~#1} 
\newcommand{\rpb}[1][$R_{\oplus}$]   {$3.41 _{ - 0.12 } ^ { + 0.14 }$~#1} 
 
\newcommand{\eb}[1][ ]   {$0.14 _{ - 0.10 } ^ { + 0.25 }$~#1} 
 
\newcommand{\ib}[1][deg]   {$88.85 _{ - 0.73 } ^ { + 0.73 }$~#1} 
 
\newcommand{\ab}[1][AU]   {$0.1053 _{ - 0.0031 } ^ { + 0.003 }$~#1}

\newcommand{\insolationb}[1][${\rm F_{\oplus}}$]   {$255.7 _{ - 19.7 } ^ { + 21.6 }$~#1} 
 
\newcommand{\denstrb}[1][${\rm g\,cm^{-3}}$]   {$0.568 _{ - 0.043 } ^ { + 0.042 }$~#1}

\newcommand{\ttotb}[1][hours]   {$5.53 _{ - 0.11 } ^ { + 0.11 }$~#1}

\newcommand{\denpb}[1][${\rm g\,cm^{-3}}$]   {$1.58 _{ - 0.83 } ^ { + 0.96 }$~#1}

\newcommand{\Tzeroc}[1][days]   {$2002.7708 _{ - 0.0011 } ^ { + 0.0011 }$~#1} 
\newcommand{\Pc}[1][days]   {$24.38 _{ - 3.4 } ^ { + 6.23 }$~#1} 
\newcommand{\esinc}[1][ ]   {$0.05 _{ - 0.21 } ^ { + 0.19 }$~#1} 
\newcommand{\ecosc}[1][ ]   {$0.04 _{ - 0.37 } ^ { + 0.38 }$~#1} 
\newcommand{\bc}[1][ ]   {$0.49 _{ - 0.11 } ^ { + 0.10 }$~#1} 
\newcommand{\rrc}[1][ ]   {$0.03764 _{ - 0.00074 } ^ { + 0.00069 }$~#1}

\newcommand{\rpc}[1][$R_{\oplus}$]   {$5.83 _{ - 0.14 } ^ { + 0.14 }$~#1} 
 
\newcommand{\ec}[1][ ]   {$0.115 _{ - 0.08 } ^ { + 0.173 }$~#1} 
 
\newcommand{\ic}[1][deg]   {$88.89 _{ - 0.15 } ^ { + 0.18 }$~#1}

\newcommand{\ttotc}[1][hours]   {$6.359 _{ - 0.071 } ^ { + 0.087 }$~#1}

\newcommand{\qone}[1][]   {$0.183 _{ - 0.09 } ^ { + 0.156 }$~#1} 
\newcommand{\qtwo}[1][]   {$0.47 _{ - 0.31 } ^ { + 0.35 }$~#1}

\newcommand{\EXPRES}[1][${\rm km\,s^{-1}}$]   {$0.006 _{ - 0.0021 } ^ { + 0.0021 }$~#1} 
\newcommand{\HARPSN}[1][${\rm km\,s^{-1}}$]   {$0.0007 _{ - 0.0012 } ^ { + 0.0013 }$~#1} 
\newcommand{\jEXPRES}[1][${\rm m\,s^{-1}}$]   {$1.06 _{ - 0.82 } ^ { + 1.88 }$~#1} 
\newcommand{\jHARPSN}[1][${\rm m\,s^{-1}}$]   {$3.02 _{ - 1.27 } ^ { + 1.47 }$~#1} 
\newcommand{\jtr}[1][]   {$39 _{ - 27 } ^ { + 35 }$~#1} 

\newcommand{\mpcup}[1][$M_{\oplus}$]{$27.5$~#1}

\title[\target]{Planet Hunters TESS III: two transiting planets around the bright G~dwarf HD 152843}

\author[Eisner et al.]
{N. L. Eisner,$^{1}$\thanks{E-mail: nora.eisner@new.ox.ac.uk}
B. A. Nicholson,$^{1,2}$
O. Barrag\'an,$^{1}$
S. Aigrain,$^{1}$
C. Lintott,$^{1}$
L. Kaye,$^{1}$
\newauthor
B. Klein,$^{1}$
G. Miller,$^{1}$
J. Taylor,$^{1}$
N. Zicher,$^{1}$
L. A. Buchhave,$^{3}$
D. A. Caldwell,$^{4}$
\newauthor
J. Horner,$^{2}$
J. Llama,$^{5}$
A. Mortier,$^{6,7}$
V. M. Rajpaul,$^{6}$
K. Stassun,$^{8}$
A. Sporer,$^{9}$
\newauthor
A. Tkachenko,$^{10}$
J. M. Jenkins,$^{11}$
D. Latham,$^{12}$
G. Ricker,$^{9}$
S. Seager,$^{9,13,14}$
J. Winn,$^{15}$
\newauthor
S. Alhassan,$^{16}$
E. M. L. Baeten,$^{16}$
S. J. Bean,$^{16}$
D. M. Bundy,$^{16}$
V. Efremov,$^{16}$
\newauthor
R. Ferstenou,$^{16}$
B. L. Goodwin,$^{16}$
M. Hof,$^{16}$
T. Hoffman,$^{16}$
A. Hubert,$^{16}$
L. Lau,$^{16}$
\newauthor
S. Lee,$^{16}$
D. Maetschke,$^{16}$
K. Peltsch$^{16,17}$
C. Rubio-Alfaro$^{16}$
and G. M. Wilson$^{16}$
\\ 
$^{1}$Department of Physics, University of Oxford, Keble Road, Oxford OX1 3RH, UK\\
$^{2}$Centre for Astrophysics, University of Southern Queensland, Toowoomba, Queensland 4350, Australia\\
$^{3}$DTU Space, National Space Institute, Technical University of Denmark, Elektrovej 328, DK-2800 Kgs. Lyngby, Denmark \\
$^{4}$ SETI Institute 189 Bernardo Ave, Suite 200 Mountain View, CA 94043, USA \\
$^{5}$ Lowell Observatory, 1400 W. Mars Hill Rd., Flagstaff, AZ 86001, USA \\
$^{6}$Astrophysics Group, Cavendish Laboratory, University of Cambridge, J.J. Thomson Avenue, Cambridge CB3 0HE, UK \\
$^{7}$Kavli Institute for Cosmology, University of Cambridge, Madingley Road, Cambridge CB3 0HA, UK\\
$^{8}$Department of Physics and Astronomy, Vanderbilt University, Nashville, TN 37235, USA \\
$^{9}$Department of Physics and Kavli Institute for Astrophysics and Space Research, Massachusetts Institute of Technology, Cambridge, MA 02139, USA \\
$^{10}$ Institute of Astronomy, KU Leuven, Celestijnenlaan 200D, 3001 Leuven, Belgium\\
$^{11}$ NASA Ames Research Center, Moffett Field, CA 94035, USA \\
$^{12}$ Harvard-Smithsonian Center for Astrophysics, 60 Garden St., Cambridge, MA 02138, USA \\
$^{13}$Department of Earth, Atmospheric and Planetary Sciences, Massachusetts Institute of Technology, Cambridge, MA 02139, USA \\
$^{14}$Department of Aeronautics and Astronautics, MIT, 77 Massachusetts Avenue, Cambridge, MA 02139, USA \\
$^{15}$ Department of Astrophysical Sciences, Princeton University, 4 Ivy Lane, Princeton, NJ 08544, USA \\
$^{16}$ Citizen Scientist, Zooniverse c/o University of Oxford, Keble Road, Oxford OX1 3RH, UK \\
$^{17}$ School of Computer Science \& Technology, Algoma University, Sault Ste. Marie, Ontario, P6A 2G4, Canada}

\date{Accepted 2021 April 20. Received 2021 April 7; in original form 2021 March 13}

\pubyear{2021}

\begin{document}
\label{firstpage}
\pagerange{\pageref{firstpage}--\pageref{lastpage}}
\maketitle
\begin{abstract}
We report on the discovery and validation of a two-planet system around a bright (V = 8.85 mag) early G dwarf (1.43\,\Rsun, 1.15\,\Msun, TOI 2319) using data from NASA’s Transiting Exoplanet Survey Satellite (\textit{TESS}). Three transit events from two planets were detected by citizen scientists in the month-long \textit{TESS} light curve (sector 25), as part of the Planet Hunters TESS project. Modelling of the transits yields an orbital period of \Pb\ and radius of \rpb\ for the inner planet, and a period in the range 19.26--35 days and a radius of \rpc\ for the outer planet, which was only seen to transit once. Each signal was independently statistically validated, taking into consideration the \textit{TESS} light curve as well as the ground-based spectroscopic follow-up observations. Radial velocities from HARPS-N and EXPRES yield a tentative detection of planet\,b, whose mass we estimate to be \mpb, and allow us to place an upper limit of \mpcup\ (99 per cent confidence) on the mass of planet\,c. Due to the brightness of the host star and the strong likelihood of an extended H/He atmosphere on both planets, this system offers excellent prospects for atmospheric characterisation and comparative planetology. 
\end{abstract}




\section{Introduction}

Systems with multiple transiting planets offer a wealth of information for exoplanetary science. In particular they allow for comparative planetology: studying planets that have formed out of the same material, but have formed and evolved in different environments, receiving different amounts of incident flux from the host star, resulting in differing masses, radii and composition. Well characterised multi-planet systems therefore provide important model constraints that single-planet systems cannot, providing insight into planetary system architecture and evolutionary pathways, as well as informing ongoing planet population studies \citep[e.g, ][]{Tremaine2012, Dietrich2020}. 

The \textit{Kepler} mission \citep{Borucki2010} revealed that multi-planetary systems are common \citep{Latham2011}, with almost half of all \textit{Kepler} planets listed in the NASA Exoplanet Archive belonging to multi-planet systems \citep{Akeson2013}. However, the majority of the hundreds of multi-planet systems found by \textit{Kepler} are too faint to follow-up with ground-based high-resolution spectroscopy. This has resulted in most known multi-planet systems lacking well determined masses, densities, bulk compositions and atmospheric characterisation, all of which are key to helping us understand the overall planet population.

NASA's Transiting Exoplanet Survey Satellite \citep[TESS; ][]{ricker15}, however, targets stars that are on average a 30-100 times brighter than those observed by the \textit{Kepler} mission, thus allowing us to follow up and constrain the properties of systems that were previously inaccessible. \tess\ has already discovered tens of previously unknown, multi-planet systems \citep[e.g.,  ][]{2019Gandolfi, quinn2019, 2019Dragomir, Gilbert2020, 2020Mann, 2020Fridlund, Carleo2020, Leleu2021}.

Detecting transiting multi-planet systems with longer-period planets is challenging due to the reduced transit probability of those planets, as well as the challenges associated with detecting planets showing single transits using automated detection algorithms. For this reason, alternative methods are often used to identify longer-period, single transit candidates, such as machine learning \citep[e.g.,][]{Pearson2018, Zucker2018}, or visual vetting with the help of citizen science \citep{eisner2020method,fischer12}. 

Furthermore, verifying the planetary nature of single transit objects is challenging, as the lack of a known orbital period complicates follow-up efforts. However, this is made easier in the situation of multi-planet systems. \cite{Latham2011} and \cite{Lissauer2012} independently showed that systems with multiple planet candidates are statistically less likely to be false positives, compared to single-planet systems. This is helpful to consider in following up single-transit, longer-period planets with closer companions which are themselves more easily verifiable as true planetary companions. 

Despite the large number of exoplanet discoveries made by \tess\ and \textit{Kepler}, systems with more than one transiting planet around stars brighter than $V \sim 10$ \citep[the typical magnitude required for atmospheric follow-up, e.g., ][]{Fortenbach2020} containing planets with measured masses remain exceedingly rare. As of \textcolor{black}{April} 2021, there are only 17 transiting planets (in 12 systems) with mass measurements better than 50 per cent precision around stars with V < 10 listed in the NASA Exoplanet Archive \citep{Akeson2013}. \textcolor{black}{A list of these systems and their corresponding parameters can be found in  Appendix~\ref{appendixA}}. Significant observing resources have been, and continue to be, devoted to each of them. 

In this paper we present a new multi-planet system, with the discovery of two planets orbiting around \target. These candidates were initially identified in \tess\ Sector 25 by citizen scientists taking part in the Planet Hunters TESS project \citep{eisner2020method}. In Section~\ref{sec:data} we outline the discovery of the candidates and the vetting tests carried out based on the \tess\ photometric light curve. In Section~\ref{sec:RecSpec} we discuss the spectroscopic data obtained with HARPS-N and EXPRES and in Section~\ref{sec:analysis} we discuss the joint photometric and spectroscopic data analysis. Finally, the results are discussed in Section~\ref{sec:results} and the conclusions presented in Section~\ref{sec:conclusions}.

\section{TESS Photometry}
\label{sec:data}

\target\ was observed by \tess\ only in Sector 25 of the primary mission. The spacecraft obtained images at a cadence of two-seconds, which were combined on board into two-minute cadence data products. These were processed and reduced by the Science Processing Operations Center \citep[SPOC; ][]{jenkinsSPOC2016}. Throughout this work we use the pre-search data conditioning (PDC) light curve from the SPOC pipeline, as shown in Figure~\ref{fig:full_LC}. The data gap seen in the centre of the full light curve corresponds to the time taken ($\sim$~1 day) for the spacecraft to send the data to Earth and re-orient itself. The black dashed lines at the bottom of the figure indicate the times of the periodic momentum dumps caused by the firing of the thrusters as the spacecraft \textcolor{black}{adjusts} the spin rate of the reaction wheels approximately every 5.5 days.  

\subsection{Discovery of \planetb\ and \planetc}
\label{subsec:discovery}

\begin{figure*}
    \centering
    \includegraphics[width=1\textwidth]{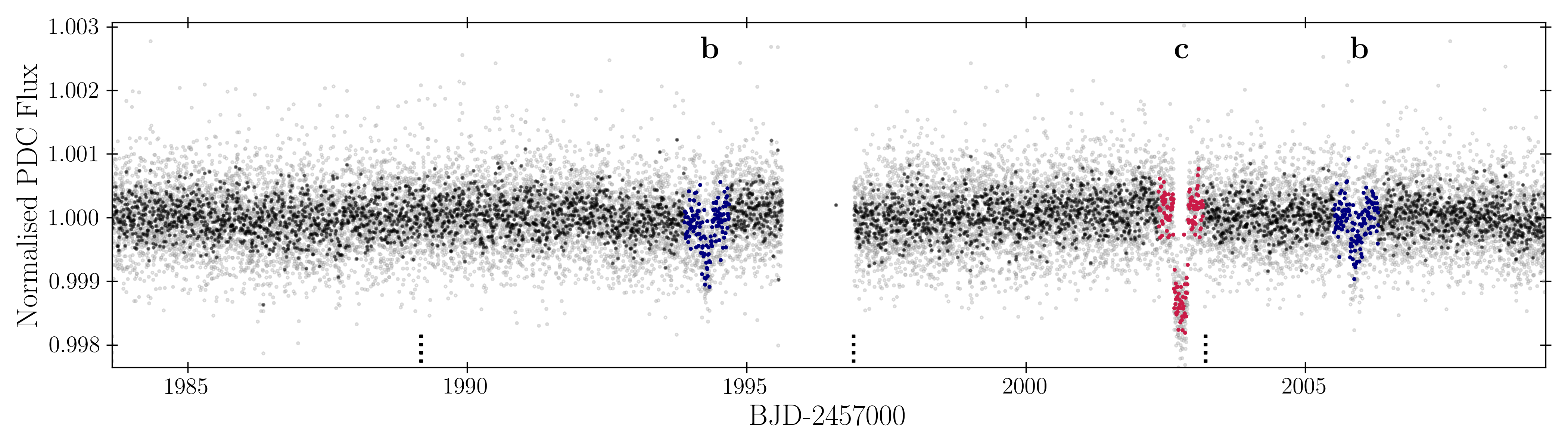}
    \caption{Flux time series for \target\ vs TESS Julian day (BJD-2457000.0) for Sectors 25. The light grey points show the short cadence data with a 2 minute sampling, whilst the black points are 10 minute averages. The dashed vertical lines at the bottom of the figure show the times of the \tess\ momentum dumps. The transit events are shown in blue and pink, corresponding to the inner and outer planet candidates.} 
    \label{fig:full_LC}
\end{figure*}

The light curve shown in Figure~\ref{fig:full_LC} exhibits three transit events belonging to different transiting planets, with \planetb\ shown in blue and \planetc\ shown in pink. The first transit event of \planetb\ (T$_{\mathrm{BJD- 2457000}}\sim$1994.28\,d) and the single transit event of \planetc\ (T$_{\mathrm{BJD- 2457000}}\sim$2002.77\,d) were flagged as a single Threshold Crossing Event (TCE) by the SPOC pipeline, as two events caused by the same `object'. However, due to the different depths of these two transits the TCE was not promoted to TESS Object of Interest (TOI) status, due to the assumption that the two events correspond to the primary and secondary eclipses of an eclipsing binary. The second transit event of \planetb\ was not flagged by the pipeline. 

All three transit events were identified by the Planet Hunters TESS (PHT) citizen science project \citep{eisner2020method}. PHT, which is hosted by the Zooniverse platform \citep{lintott08, lintott11}, harnesses the power of over 25 thousand registered citizen scientists who visually vet all of the \tess\ two-minute cadence light curves in search for transit events that were ignored or missed by the main transit detection pipeline and other teams of professional astronomers. The light curve of \target\ was seen by 15 citizen scientists, 12 of whom identified all three transit events, and 3 who identified only two out of the three events. The target was initially brought to the attention of the PHT research team via the PHT discussion forum~\footnote{\url{https://www.zooniverse.org/projects/nora-dot-eisner/planet-hunters-tess/talk/2112/1552434?comment=2520798}}. We uploaded both planet candidates to the Exoplanet Follow-up Observing Program for TESS (ExoFOP-TESS) site on 2020-08-07 as a community TESS Object of Interest (cTOI). The inner planet has since been promoted to the priority 1 (1 = highest priority, 5 = lowest priority) candidate TOI 2319.01.

\subsection{Excluding false positive scenarios}

Astrophysical and instrumental false positives are common in the \tess\ data, in particular due to the large (21 arcsec pix$^{-1}$) pixel scale. We used the publicly available Lightcurve Analysis Tool for Transiting Exoplanets \citep[{\sc latte}; ][]{LATTE2020} in order to perform standard diagnostic tests that help to rule out false positive scenarios including background eclipsing binaries, systematic effects, and background events such as asteroids passing through the field of view. For a full description of the diagnostic tests we refer the reader to \cite{LATTE2020}, however in brief the tests include: 

\begin{enumerate}
    \item Checking that the transit events do not coincide with the times of the periodic momentum dumps.
    \item Checking that the x and y centroid positions are smoothly varying with time in the vicinity of the transit events.
    \item Examining light curves of the five nearest two-minute cadence \tess\ stars to check for systematic effects.
    \item Examining light curves extracted for each pixel surrounding the target in order to ensure that the signal is not the result of a background eclipsing binary, a background event or caused by systematics.
    \item Checking that there are no spurious signals, such as sudden jumps or strong variations, in the background flux.
    \item Comparing transit shapes and depths when extracted with different aperture sizes.
    \item {\color{black} Comparing} between the average in-transit and average out-of-transit flux, as well as the difference between them.
    \item Checking the location of nearby stars brighter than V-band magnitude 15 as queried from the Gaia Data Release 2 catalog \citep{gaia2018gaia}.
    \item Performing the box-Least-Squares fit to search for additional signals.
\end{enumerate}

Tests (i) to (iv) enabled us to rule out events caused by systematic effects due to the satellite or instrument, and tests (iii) to (viii) increased our confidence that the signals are not caused by astrophysical false positives, such as blends where the photometric aperture of a bright target contains a faint eclipsing binary. 

\begin{figure}
    \centering
    \includegraphics[width=0.3\textwidth]{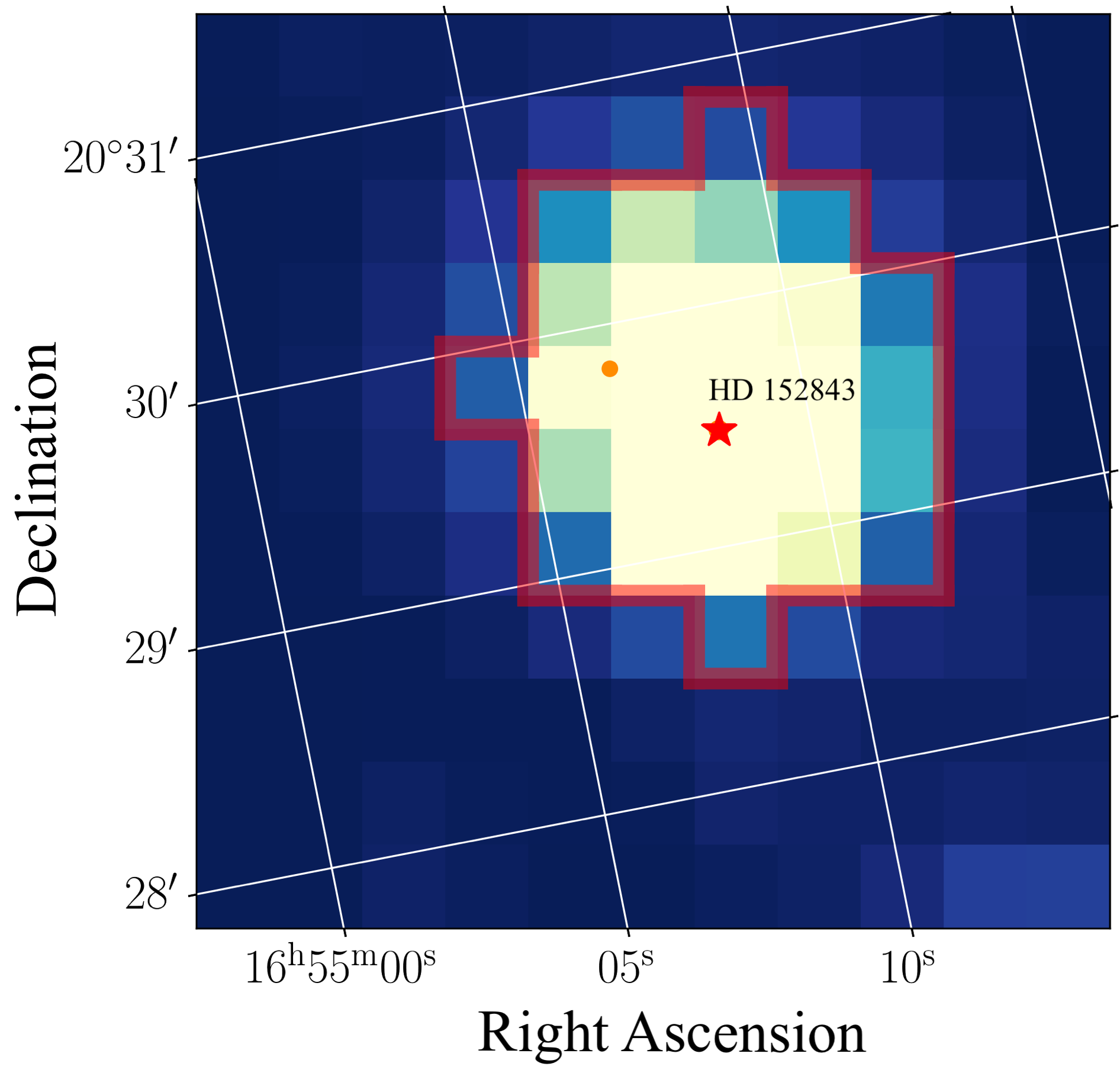}
    \caption{The median \tess\ image around \target. The aperture used to extract the light curve is shown by the red outline and the orange dot depicts the location of the only star brighter than V = 15 within 110 arcseconds of the target (red star), as queried by Gaia DR2 \citep{gaia2018gaia}. This nearby star (V = 14.4) is located at an angular separation of $\sim$ 31.3 arcsec.} 
    \label{fig:nearby_stars}
\end{figure}

As blends are common in the \tess\ data, we searched for nearby Gaia Data Release 2 catalog stars \citep{gaiadr2} within 110 arcseconds of the target, and found there to only be a single star with a V-band magnitude brighter than 15, as shown by the orange circle in Figure~\ref{fig:nearby_stars}, where the red star shows \target\ and the red outline highlights the aperture used to extract the light curve.

In order to rule out this nearby star as the cause of the transit events, we calculated the magnitude difference between \target\ and the faintest companion star that could plausibly be responsible for the observed transit shapes and depths. Following the methodology outlined by \cite{Vanderburg2019} and the transit parameters derived using \texttt{pyaneti} (see Section~\ref{subsec:modelling}) we show that the maximum magnitude difference between the target star and a possible background contaminant is 1.5 magnitude in the V band. This allows us to confidently conclude that the 14.4 magnitude star (5.6 magnitude fainter than \target), located at an angular separation of $\sim$~31.3\,", is not responsible for either of the planetary signals.


\renewcommand{\arraystretch}{1}
\begin{table*}
\centering
  \caption{Stellar parameters. \label{tab:star}}  
  \begin{tabular}{lcc}
  \hline
  \hline
  Parameter & Value & Source  \\
  \hline
  \multicolumn{3}{l}{\bf Identifiers} \\
  HD &  152843     &       \\
  TOI &  2319      &       \\
  TIC  & 349488688 & \cite{Stassun19} \\
  Gaia DR2 & 4564566554995619072 &  \textit{Gaia} \textcolor{black}{eDR3}$^{(\mathrm{a})}$ \\
  2MASS & J16550834+2029287 & 2MASS$^{(\mathrm{b})}$ \\
  \hline
  
  \multicolumn{3}{l}{\bf Astrometry} \\
  $\alpha_{\rm J2000}$ & \textcolor{black}{16:55:08.373}  &   \textit{Gaia} \textcolor{black}{eDR3}$^{(\mathrm{a})}$ \\
  $\delta_{\rm J2000}$ & \textcolor{black}{20:29:29.509}  &  \textit{Gaia} \textcolor{black}{eDR3}$^{(\mathrm{a})}$ \\
  Distance (pc) & 107.898  $\pm$ 0.317 &  \cite{2018Bailer} \\
  $\pi$ (mas) &  \textcolor{black}{9.161 $\pm$ 0.015} & \textit{Gaia} \textcolor{black}{eDR3}$^{(\mathrm{a})}$ \\
  \hline
  \multicolumn{3}{l}{\bf Photometry} \\
  B  &  $ 9.380 \pm 0.020  $ & \textit{Tycho-2}$^{(\mathrm{c})}$\\
  V  &  $ 8.850 \pm 0.010  $ & \textit{Tycho-2}$^{(\mathrm{c})}$\\
  J  &  $ 7.896 \pm 0.018 $ & 2MASS$^{(\mathrm{b})}$\\
  H  &  $ 7.655 \pm 0.016 $ & 2MASS$^{(\mathrm{b})}$\\
  K  &  $ 7.629 \pm 0.020 $ & 2MASS$^{(\mathrm{b})}$\\
  W1 &  $ 7.563 \pm 0.031 $ & WISE$^{(\mathrm{d})}$ \\
  W2 &  $ 7.594 \pm 0.020 $ & WISE$^{(\mathrm{d})}$\\
  W3 &  $ 7.607 \pm 0.019 $ & WISE$^{(\mathrm{d})}$\\
  \hline
  \multicolumn{3}{l}{\bf Physical Properties} \\
    Spectral Type &  G0  &  \\
    Effective Temperature $\mathrm{T_{eff}}$ (K)  & \textcolor{black}{$6310 \pm 100 $ }                & This work \\
    Surface gravity $\log g_\star$ (cgs)          & \textcolor{black}{$4.19 \pm 0.03 $ }                & This work\\ 
    \vsini (\kms)                                 & \textcolor{black}{$8.38 \pm 0.50$  }                  & This work  \\
    $[M/H]$ (dex)                                 & \textcolor{black}{$-0.22 \pm 0.08$ }                & This work \\
    $[Fe/H]$ (dex)                                & \textcolor{black}{$-0.16 \pm 0.05$ }                & This work \\
    $v_{\rm mic}$ (\kms)                          & \textcolor{black}{$1.66 \pm 0.13$  }                 & This work \\
    $v_{\rm mac}$ (\kms)                          & \textcolor{black}{2}                           & \cite{Bruntt2010} \\
    Stellar mass $M_{\star}$ ($M_\odot$)          & \textcolor{black}{$1.15 \pm 0.04 $ }         & This work \\
    Stellar radius $R_{\star}$ ($R_\odot$)        & \textcolor{black}{$1.43 \pm 0.02 $ }        & This work \\
    Stellar density $\rho_\star$ ($\rho_\odot$)   & \textcolor{black}{$0.40 \pm 0.03 $ }         & This work \\  
    Star age (Gyr)                                & \textcolor{black}{$3.97 \pm 0.75$  }          & This work \\   
    \hline
  \end{tabular}
     \begin{tablenotes}\footnotesize
  \item \textit{Note} -- $^{(\mathrm{a})}$ \textit{Gaia} \textcolor{black}{early Data Release 3} \citep[eDR3; ][]{gaiaedr3}. $^{(\mathrm{b})}$ Two-micron All Sky Survey \citep[2MASS; ][]{2MASS2003}. $^{(\mathrm{c})}$ \textit{Tycho}-2 catalog \citep{Hog2000}.  $^{(\mathrm{d})}$ Wide-field Infrared Survey Explorer catalog \citep[WISE; ][]{Cutri2013}
\end{tablenotes}
\end{table*}



\subsection{Limits on additional planets}
\label{sec:injectiontest}

We quantify the detectability of additional planets in the \textit{TESS} light curve using a transit injection and recovery test \citep[e.g., ][]{eisner2020TOI813}. In brief, we removed the known transit events prior to injecting synthetic signals into the PDC \tess\ light curve. The injected signals were generated using the {\sc batman} package \citep{Kreidberg15}, with planet radii ranging from 1 to 12\,R$_\oplus$ and periods ranging from 3 to 24~days, both sampled at random from a log-uniform  distribution. The impact parameter and eccentricity were assumed to be zero throughout and we used a quadratic limb-darkening law with $q1$ and $q2$ of 0.16 and 0.59, respectively, as taken from Table~15 in \cite{Claret2017} using the stellar parameters given in Table~\ref{tab:star}. Once the signals were injected, we used an iterative non-linear filter \citep{Aigrain04} to estimate and subtract residual systematics on timescales > 1.7 days.

We simulated and injected a total of 750,000 transit events. The Box Least Squares \citep[BLS; ][]{Kovacs2002} algorithm was then used to try to recover the injected signals. The BLS search sampled a frequency grid that was evenly-spaced from 0.01 to 1 day$^{-1}$. For each simulation, we recorded the period and orbital phase corresponding to the highest peak in the BLS periodogram. If the recovered orbital period and phase agreed to within 1~ per cent of the injected period, the signal was deemed to be correctly identified. The completeness, assessed in radius and period bins with width of 0.25\,R$_\oplus$ and 0.75~d respectively, was then taken to be the fraction of correctly identified transit signals. 

\begin{figure}
    \centering
    \includegraphics[width=0.52\textwidth]{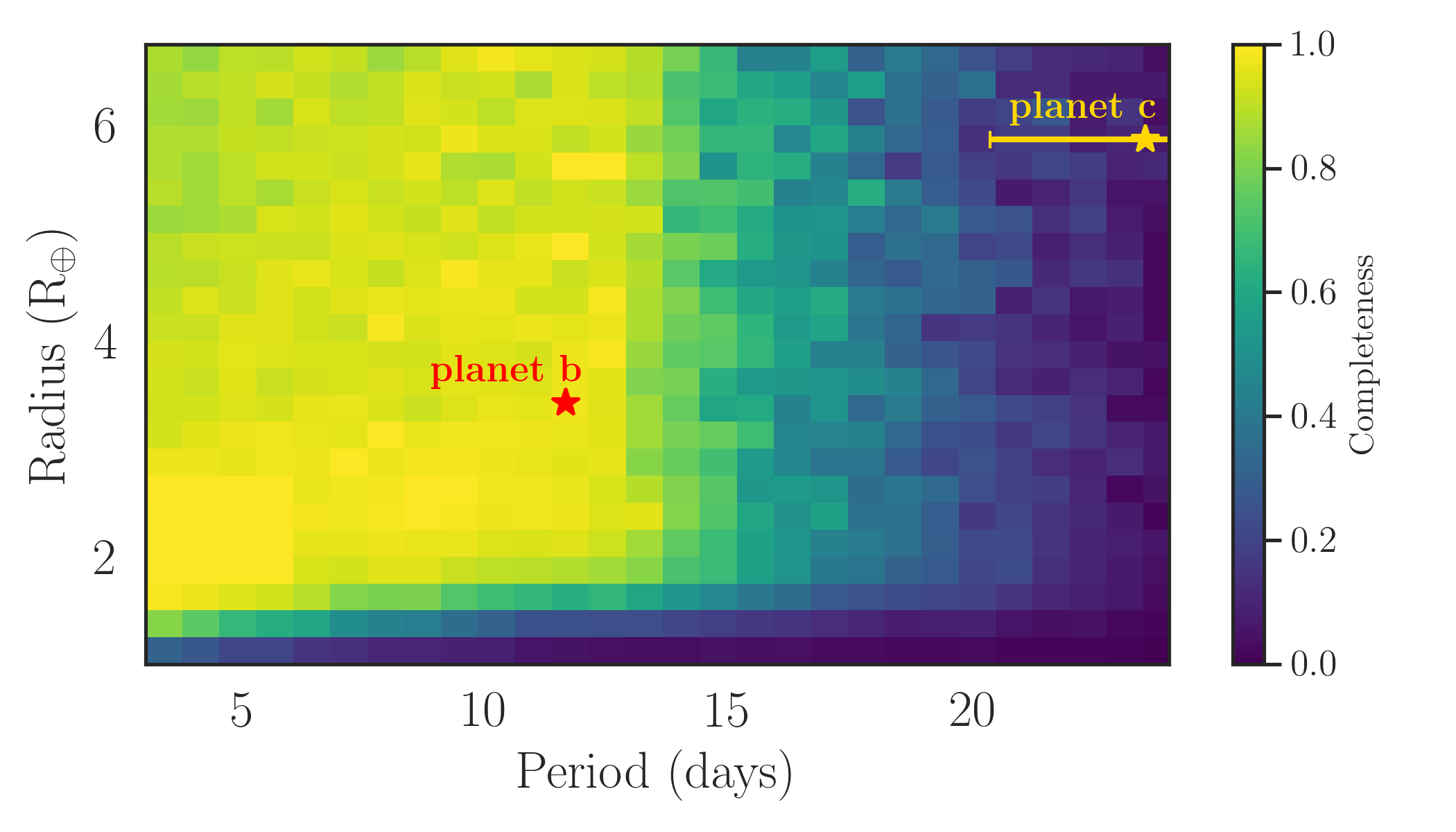}
    \caption{The recovery completeness of injected transit signals
into the light curve of \target\ as a function of the radius and orbital period. The signals were recovered using a BLS search. The properties of \planetb\ and \planetc\ are shown by the red and yellow star respectively.}
    \label{fig:inj_rec}
\end{figure}

The results, shown in Figure~\ref{fig:inj_rec}, highlight, as expected, that the automated BLS search is strongly biased towards detecting shorter period planets that transit multiple times in the light curve. The limited duration of the \tess\ observations of $\sim$ 27~d, interrupted by a 1.3 day data gap, results in a sharp decline in completeness for periods longer than around 13 days. For planets greater than 2 \REarth we recover 94 per cent of signals with periods between 12 and 13 days and 78 per cent of signals with periods between 14 and 15 days. The completeness for the parameters of planet\,b is close to 100~ per cent, while the completeness for the parameters of planet\,c is close to 0 per cent due to the fact that there is only one transit within the available \tess\ light curve. We caution that the simulated signals were injected into the PDC light curve, which has already undergone detrending and systematics corrections by the SPOC pipeline. The presented recovery rates are, therefore, systematically higher than one might otherwise expect if the signals had been injected into the raw light curve \citep[e.g., ][]{2020Lienhard}. Overall, this analysis highlights the difficulties associated with detecting longer-period planets using automated algorithms, and demonstrates a need for alternative detection methods such as citizen science.

\section{Spectroscopic data}
\label{sec:RecSpec}

\subsection{Reconnaissance spectra} 
\label{subsec: nres}

We made use of the Las Cumbres Observatory (LCO) telescopes with the Network of Robotic Echelle Spectrographs \citep[NRES;][]{LCO2013}. This fibre-fed spectrograph, mounted on a 1.0-m telescope, has a resolution of R = 53,000 and a wavelength coverage of 380 to 860 nm. We obtained two spectra of \target\ on the 15th and 22nd August 2020 with per pixel signal to noise ratios (SNR) of 38 and 25 at 520 nm, respectively. The two spectra gave radial velocity estimates of $9.7\pm0.2$ km/s and $9.6\pm0.7$ km/s, which are consistent within their uncertainties, and thus allowed us to rule out the possibility that the transit events are caused by an eclipsing binary.

\subsection{High-resolution spectra} 
\label{subsec: high-res}


We acquired high-resolution (R\,$\approx$\,115\,000) spectra with the High Accuracy Radial velocity Planet Searcher in the Northern hemisphere \citep[HARPS-N;][]{Cosentino2012,Cosentino2014} spectrograph mounted at the 3.6-m Telescopio Nazionale Galileo in La Palma, Spain, via Director's Discretionary Time (program ID A41DDT4). We obtained 18 spectra between 2020 September 5 and November 11 (mean SNR $\sim$ 89 at at 550 nm). Each spectrum has simultaneous wavelength calibration with a Fabry-Perot etalon and was reduced via the standard HARPS Data Reduction Software \citep[DRS; ][]{1996Baranne} using a G2 spectral template (mean RV uncertainty $\sim$ 4.2 \ms). Additionally, we extracted the HARPS-N radial velocity (RV) measurements using the {\sc TERRA} pipeline \citep{2012TERRA}, which uses a template-matching approach based on a template generated by stacking all of the spectra. The results extracted using DRS and {\sc TERRA} have comparable uncertainties, with a slightly larger root-mean-square scatter in the {\sc TERRA} extracted data. Around 71 per cent of the DRS/TERRA RVs agree within 1$\sigma$ and around 82 per cent agree within 2$\sigma$. For the remainder of our analysis we used the data extracted with the DRS. 

\textcolor{black}{We derived the $\log R'_{\rm HK}$ values for the HARPS-N spectra with SNR > 100 using the calibrations of \cite{1984Noyes}, and found the values to range from -4.96 to -4.94 with a mean value of -4.95. This low value suggests that \target\ is a quiet star. We also note that there is no correlation between the $\log R'_{\rm HK}$ values and the radial velocities.}

\textcolor{black}{In addition to the HARPS-N observations} we obtained 22 spectra between 9 September and 10 October 2020 using the high-resolution (R\,$\approx$\,150\,000) EXtreme PREcision Spectrometer \citep[EXPRES;][]{Jurgenson2016, Petersburg2020, Blackman2020} mounted on the 4.3-m Lowell Discovery Telescope \citep[LDT; ][]{levine2012}, USA. Each spectrum was calibrated using a Thorium Argon lamp and a stabilized Laser Frequency Comb and the RVs were extracted using the EXPRES analysis pipeline \citep[for detail see ][]{Petersburg2020}. Due to poor seeing and high airmass, 12 of those spectra (with SNR < 25 at 550 nm) were not used for further analysis. The mean SNR and mean RV uncertainty of the used spectra are $\sim$ 82 and $\sim$ 9.5 \ms, respectively. All HARPS-N and EXPRES RV measurements are listed in Table~\ref{tab:RVs}.

\renewcommand{\arraystretch}{1}
\begin{table*}
\centering
  \caption{Radial velocity measurements. \label{tab:RVs}}  
  \begin{tabular}{ccccc}
  \hline
  \hline
	Time			& RV			&	$\sigma_{RV}$	&	SNR		&	Source	\\
	(BJD-2457000)	& (\ms)	  		&	(\ms)			&			&		    \\
	\hline
	2098.3521		& 4.2460	  	&	1.7400			&	155.1 	&	HARPS-N	\\
	2101.6407		& -3.4170	  	&	12.1000			&	19.0 	&	EXPRES*	\\
	2101.6553		& 19.1100	  	&	13.8290			&	16.0 	&	EXPRES*	\\
	2101.6701		& -27.3050  	&	14.6120			&	14.0 	&	EXPRES*	\\
	2101.6849		& -26.5790		&	14.0830			&	14.0 	&	EXPRES*	\\
	2102.3412		& -4.4874		&	2.3570			&	117.5 	&	HARPS-N	 \\
	2102.6207		& 11.1370		&	11.9490			&	20.0 	&	EXPRES*	\\
	2102.6351		& 10.9350		&	10.7270			&	21.0 	&	EXPRES*	\\
	2102.6519		& 15.8470		&	12.4080			&	20.0 	&	EXPRES*	\\
	2102.6656		& -0.8630		&	11.5030			&	23.0 	&	EXPRES*	\\
	2102.6843		& 26.9160		&	11.7640			&	22.0 	&	EXPRES*	\\
	2102.6999		& -24.1910		&	11.2430			&	22.0 	&	EXPRES*	\\
	2102.7140		& -15.1840		&	12.8060			&	18.0 	&	EXPRES*	\\
	2102.7312		& -43.8920		&	14.5010			&	13.0 	&	EXPRES*	\\
	2104.3651		& -8.9263		&	18.0372			&	19.9 	&	HARPS-N*\\
	2110.3253		& -3.5140		&	3.2192			&	86.6 	&	HARPS-N	\\
	2111.3788		& 1.8856		&	2.7648			&	99.8 	&	HARPS-N	\\
	2117.3242		& 5.3618		&	2.9210			&	95.2 	&	HARPS-N	\\
	2119.3255		& 5.5608		&	3.2501			&	78.5 	&	HARPS-N	\\
	2120.3307		& 0.2539		&	2.2039			&	126.2 	&	HARPS-N	\\
	2120.4134		& 3.4885		&	3.3055			&	85.9 	&	HARPS-N	\\
	2120.6143		& 5.4070		&	4.9410			&	95.0 	&	EXPRES	\\
	2123.5929		& -0.2250		&	5.3170			&	81.0 	&	EXPRES	\\
	2123.6069		& 0.1490		&	4.9690			&	83.0 	&	EXPRES	\\
	2125.3192		& -1.6649		&	2.4766			&	110.9 	&	HARPS-N	\\
	2126.3165		& -6.6282		&	4.4202			&	64.7 	&	HARPS-N	\\
	2126.5970		& -1.0650		&	8.9100			&	41.0 	&	EXPRES	\\
	2126.6118		& 11.0790		&	6.8700			&	57.0 	&	EXPRES	\\
	2127.3185		& 4.9884		&	3.9503			&	71.8 	&	HARPS-N	\\
	2128.3180		& 9.1121		&	3.1712			&	87.2 	&	HARPS-N	\\
	2129.5838		& 5.0890		&	5.1920			&	92.0 	&	EXPRES	\\
	2129.5967		& 13.1530		&	5.8300			&	64.0 	&	EXPRES	\\
	2130.3156		& 1.4459		&	6.3500			&	45.3 	&	HARPS-N	\\
	2130.5850		& 9.9870		&	5.1790			&	90.0 	&	EXPRES	\\
	2132.5928		& 8.5890		&	4.4480			&	114.0 	&	EXPRES	\\
	2132.6078		& 5.3240		&	4.8730			&	110.0 	&	EXPRES	\\
	2152.2939		& 1.4524		&	2.5161			&	112.1 	&	HARPS-N	\\
	2153.2902		& -4.2017		&	2.4502			&	115.7 	&	HARPS-N	\\
	2154.2902		& 3.2138		&	3.6046			&	80.8 	&	HARPS-N	\\
	2155.2908		& -11.5868		&	7.1983			&	44.0 	&	HARPS-N	\\
    \hline
   \noalign{\smallskip}
  \end{tabular}
  \begin{tablenotes}\footnotesize
  \item \textit{Note} -- * indicates that the spectrum was not used for further analysis due to low signal to noise (SNR < 45). The SNRs are calculated at 550 nm. 
\end{tablenotes}
\end{table*}


\section{Data analysis}
\label{sec:analysis}

\subsection{Stellar atmospheric parameters}
\label{subsec:stellar}

The fundamental stellar parameters of \target, \textcolor{black}{namely} the effective temperature ($T_{\mathrm{eff}}$), surface gravity ($\log g$), metallicity ([M/H]), projected rotational velocity (\textit{vsin i}), and microturbulent velocity ($\xi_{t}$), were extracted using three independent methods: \textcolor{black}{ARES+MOOG \footnote{ARESv2:  \url{http://www.astro.up.pt/~sousasag/ares/}; MOOG 2017: \url{http://ww w.as.utexas.edu/~chris/moog.html}}, Grid Search in Stellar Parameters ({\sc gssp}) \footnote{GSSP: \url{https://fys.kuleuven.be/ster/meetings/binary-2015/gssp-software-package}}, and  Stellar Parameter Classification ({\sc SPC}).}

The ARES+MOOG method derives stellar atmospheric parameters using a curve-of-growth method based on the equivalent widths (EW) of the Fe I and Fe II lines \citep[for details see ][]{2014Sousa}. The EWs of the spectral lines were automatically extracted from a stacked spectrum of all of the HARPS-N data (with SNR > 45), using the {\sc Ares2} code \citep{2015Sousa}. The stacked spectrum has a SNR $\sim$ 350 at 6000 $\Angstrom$. The radiative transfer code {\sc MOOG} \citep{1973Sneden} was then used to extract the stellar parameters, assuming local thermodynamic equilibrium (LTE) and using a grid of ATLAS plane-parallel
model atmospheres \citep{kurucz1993}. The value of log g was subsequently further refined \citep{2014Mortier} and systematic and precision errors combined in quadrature. The method yields the following values: $T_{\mathrm{eff}}$ =  6348 $\pm$ 100 K, $\log g$ =  4.31 $\pm$ 0.12,  [Fe/H] = -0.16 $\pm$ 0.06, and $\xi_{t}$ = 1.82 $\pm$ 0.13 \kms. 

\begin{figure}
    \centering
    \includegraphics[width=0.49\textwidth]{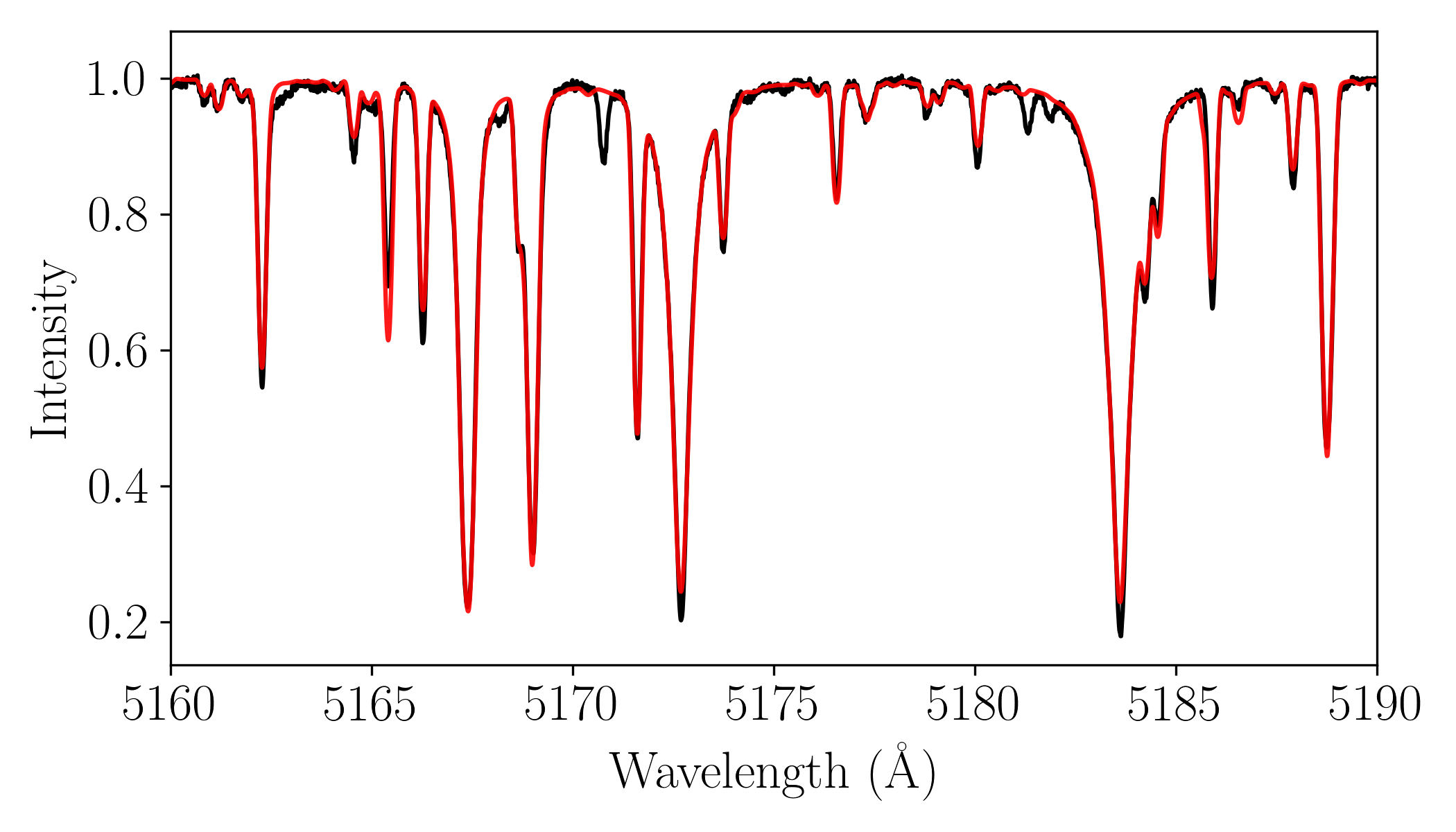}
    \caption{Section of the \textcolor{black}{stacked HARPS-N spectra with SNR > 45} (black) and the best-fit model as determined and computed with the GSSP software (red). \textcolor{black}{The parameters and abundances of this best-fit model, combined with the results from the ARES+MOOG and SPC analysis, were used to determine the stellar parameters listed in Table~\ref{tab:star}.}}
    \label{fig:spectrum}
\end{figure}

\textcolor{black}{We also used the open access {\sc gssp} code \citep[][]{tkachenko2015},} which compares the normalised observed spectrum with a grid of synthetic spectra. \textcolor{black} {A stacked spectrum of all of the HARPS-N data (with SNR > 45)} was used for this analysis. The goodness of fit of each synthetic spectrum was assessed using a ${\chi}^2$ metric. The atmospheric models used as part of this code were pre-computed using the {\sc LLmodels} software \citep{Shulyak2004} and the code assumed LTE. We independently optimised the abundances of Fe, Mg, Ti, Cr and Ni. The best-fit spectral model is shown in Figure~\ref{fig:spectrum}.

In order to determine the best-fit parameters and abundances, the ${\chi}^2$ value was recorded for each combination of parameters. The projected ${\chi}^2$ values were then fit with a  fourth order polynomial for each parameter in order to determine the global minimum, which corresponds to the value of the best-fit parameter. The uncertainties were taken as the intersection between the polynomial and the 1 $\sigma$ uncertainty limit. The following atmospheric parameters were obtained using GSSP: \textcolor{black}{$T_{\mathrm{eff}}$ = 6368 $\pm$ 100  K, log g = 4.16 $\pm$ 0.10, [M/H] = -0.17 $\pm$ 0.05, [Fe/H] = -0.16 $\pm$ 0.05, $v\sin i$ = 8.56 $\pm$  0.5 \kms and $\xi_{t}$ = 1.50 $\pm$ 0.15 \kms}. We note that the derived $v\sin i$ value is not representative of the true rotational velocity of the star; instead, it represents a combined line broadening due to rotation and macroturbulence. Since we do not rely on the rotation rate of the star in our subsequent analysis, we find disentangling the effects of rotation and macroturbulent velocity to be beyond the scope of this study. 

\textcolor{black}{Finally, we used the {\sc SPC} tool \citep[for details see][]{2012Buchhave, 2014Buchhave}. Similarly to GSSP, SPC uses spectral synthesis, which was independently carried out on each HARPS-N spectrum (where SNR > 45).} We obtained the following values: $T_{\mathrm{eff}}$ = 6175 $\pm$ 50 K,  log g  = 4.15 $\pm$ 0.10,  [M/H] = -0.26 $\pm$ 0.08, and $v\sin i$ =  8.2 $\pm$ 0.5 \kms.


The values listed in Table~\ref{tab:star}, the averages of the results obtained from these three methods, were used for all subsequent analysis. Finally, we note that the spectra show almost no sign of Ca H and K re-emission, suggesting low magnetic activity. 



\subsection{SED fitting}

\begin{figure}
    \centering
    \includegraphics[width=\linewidth,trim=100 75 90 90,clip]{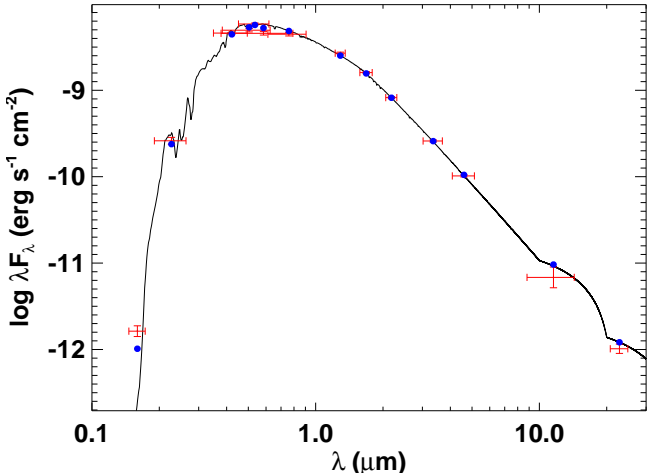}
\caption{Spectral energy distribution of \target. Red symbols represent the observed photometric measurements, where the horizontal bars represent the effective width of the passband. Blue symbols are the model fluxes from the best-fit Kurucz atmosphere model (black).  \label{fig:sed}}
\end{figure}

As an independent determination of the basic stellar parameters, we performed an analysis of the broadband spectral energy distribution (SED) of the star together with the {\it Gaia\/} DR2 parallax \citep[adjusted by $+0.08$~mas to account for the systematic offset reported by][]{StassunTorres18}, in order to determine an empirical measurement of the stellar radius, following the procedures described in \citet{Stassun2016,Stassun2017,Stassun2018}. We pulled the $B_T V_T$ magnitudes from {\it Tycho-2}, the $JHK_S$ magnitudes from {\it 2MASS}, the W1--W4 magnitudes from {\it WISE}, the $G G_{\rm BP} G_{\rm RP}$ magnitudes from {\it Gaia}, and the FUV and NUV magnitudes from {\it GALEX}. Together, the available photometry spans the full stellar SED over the wavelength range 0.15--22~$\mu$m (see Figure~\ref{fig:sed}).  

We performed a fit using Kurucz stellar atmosphere models, with $T_{\rm eff}$, [Fe/H], and $\log g$ adopted from the spectroscopic analysis. The remaining free parameter is the extinction $A_V$, which we limited to the maximum line-of-sight value from the Galactic dust maps of \citet{Schlegel98}. The resulting fit (Figure~\ref{fig:sed}) has a reduced $\chi^2$ of 1.9; the reduced $\chi^2$ improves to 1.1 if we exclude the {\it GALEX\/} FUV flux, which exhibits a modest UV excess suggestive of chromospheric activity. We find a best-fit $A_V = 0.04^{+0.05}_{-0.04}$.

Integrating the (unreddened) model SED gives the bolometric flux at Earth $F_{\rm bol} = 7.72 \pm 0.18 \times 10^{-9}$ erg~s$^{-1}$~cm$^{-2}$. Taking the $F_{\rm bol}$ and $T_{\rm eff}$ together with the {\it Gaia\/} parallax gives the stellar radius, $R_\star = 1.42 \pm 0.05$~R$_\odot$. In addition, we can estimate the stellar mass from the spectroscopic $\log g$ together with $R_\star$ from above, giving $M_\star = 1.11 \pm 0.15$~M$_\odot$, which is consistent with that empirical relations of \citet{Torres10}, giving $M_\star = 1.22 \pm 0.07$~M$_\odot$. 

Finally, we can use the star's rotation and mild UV excess (Fig.~\ref{fig:sed}) to estimate an age via empirical rotation-activity-age relations. The observed FUV excess implies a chromospheric activity of $\log R'_{\rm HK} = -4.51 \pm 0.05$ via the empirical relations of \citet{Findeisen2011}, which in turn implies a stellar rotation period of $P_{\rm rot} = 5.0 \pm 0.9$~d via the empirical relations of \citet{mamajek2008}, consistent with the upper limit $P_{\rm rot}/\sin i = 8.7$~d obtained from the spectroscopic $v\sin i$ and $R_\star$. 


\subsection{Stellar mass, radius, age, and distance}
\label{subsec:st_mass_radius}
The stellar parameters were extracted using isochrones and stellar evolutionary tracks. For this analysis, the combined \textcolor{black}{ARES+MOOG, GSSP and SPC} effective temperature and metallicity were used as inputs, along with the \textcolor{black}{Gaia eDR3} parallax, and the magnitude of the star in eight bands. All of the values used for this analysis are presented in Table~\ref{tab:star}.

For an in depth discussion of this analysis see \cite{2020Mortier}, however, in brief, this analysis made use of the {\sc isochrones} package \citep{morton2015}, using stellar models from the Dartmouth Stellar Evolution Database and from the MESA isochrones and Stellar Tracks \citep[MIST; ][]{choi2016}. We used { \sc MultiNest} \citep{2019Feroz} for the likelihood analysis and 400 live points. The analysis was run \textcolor{black}{six times}: for each of the stellar models (Dartmouth/MIST) it was run \textcolor{black}{three times} using the $T_{\mathrm{eff}}$ and metallicity from the spectroscopic analysis (Section~\ref{subsec:stellar}). The stellar values were extracted from the combined posteriors, taking the median and the 16th and 84th quantiles. The stellar mass, radius, density and age are listed in Table~\ref{tab:star}.

\subsection{Joint transit and RV modelling} \label{subsec:modelling}

The transit and RV data were jointly analysed using the open access \texttt{pyaneti} code \citep[][]{pyaneti}. In brief, \texttt{pyaneti} creates marginalised posterior distributions for different parameters by sampling the parameter space using a Markov chain Monte Carlo (MCMC) approach.
We use the limb-darkened quadratic models by \citet{Mandel2002} to fit the flattened transits. The RV data are fit with Keplerian RV models. 

We first modelled the transits. Since planet\,c transits only once, the two planets were analysed independently. For planet\,b both transits were fitted simultaneously. This allowed us to fit for transit epoch, orbital period, impact factor, scaled planet radius, and scaled semi-major axis. 

The single transit event (planet\,c) was modelled by fitting for the same parameters as for planet\,b, with the exception of the orbital period and scaled semi-major axis, as these cannot be constrained in the case of a single transit event. Instead, we obtained a possible period range of 13 to 35 days at the 99 per cent confidence interval, using the relations presented in \citet{Osborn2016} and assuming a circular orbit. These results were used to create uniform priors for all the transit model parameters, for a joint RV and transit analysis. 


All fitted parameters and priors used for the joint modeling are presented in Table~\ref{tab:parstarget}. We note that for this analysis we allow the orbits to be eccentric in order to give more flexibility to the analysis.
We sample for the stellar density $\rho_\star$, and we recover the scaled semi-major axis for each planet in the system using Kepler's third law. We use a Gaussian prior on $\rho_\star$ using \textcolor{black}{the stellar mass and radius derived in Section~\ref{subsec:st_mass_radius}.} We also note that because planet\,c only exhibits a single-transit event we use a wide uniform prior on its period, based on the results from the single-transit analysis. However, we truncated the lower period limit at 19.26 d, as a shorter orbital period would have necessarily resulted in further transit events being present within the \tess\ light curve. 



\begin{figure}
    \centering
    \includegraphics[width=0.45\textwidth]{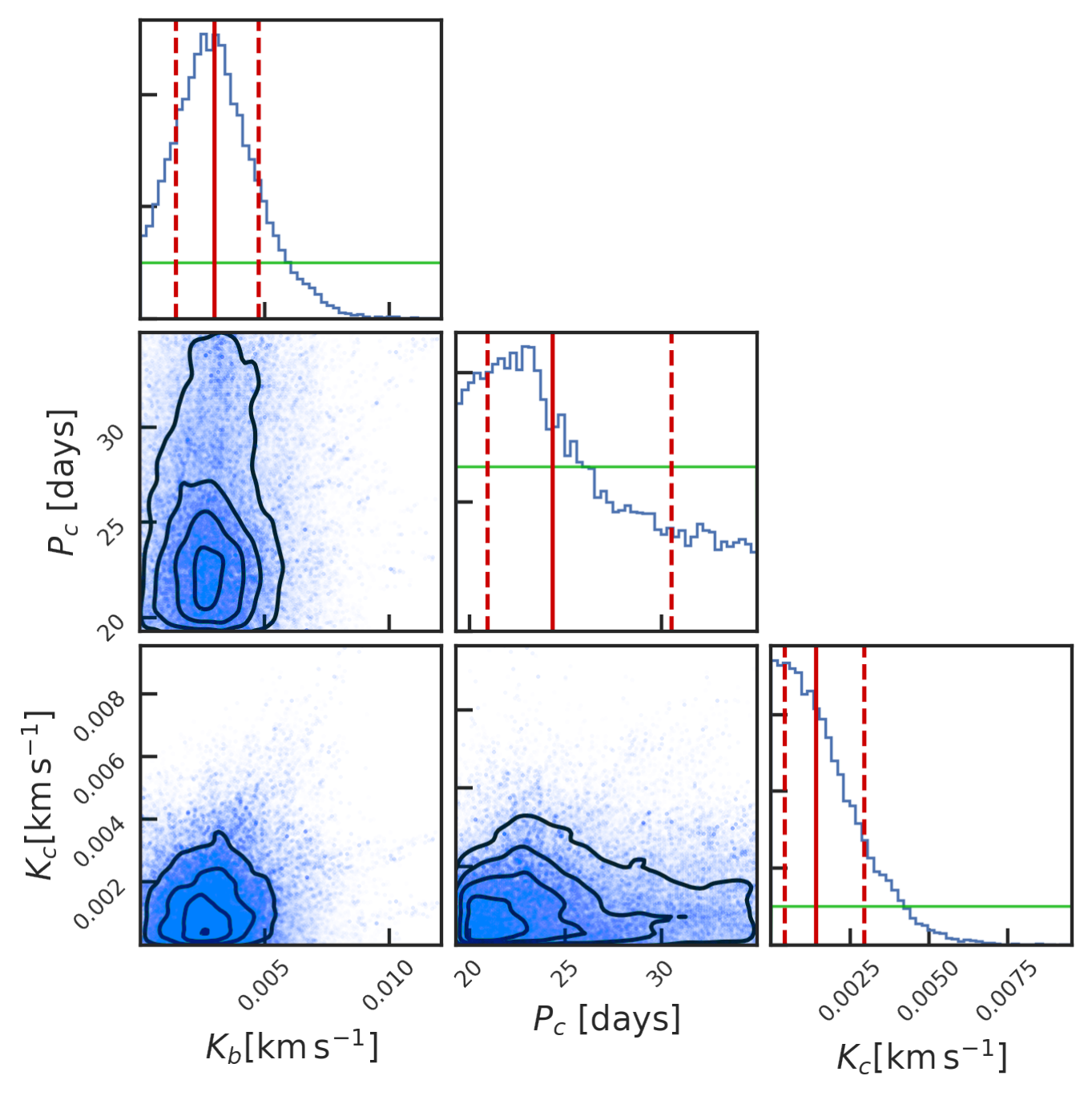}
    \caption{Corner plot for $K_b$, $P_c$, and $K_c$. First row in each column shows the posterior distribution (blue line) together with the prior shape (solid green line). Vertical solid (red) lines show the median, and vertical dashed (red) lines indicate  68.3 per cent credible intervals. The rest of sub-plots show the correlation between parameters.Transparent blue points show individual samples and solid black lines show iso-density contours. }
    \label{fig:correlations}
\end{figure}

We sampled the parameter space using an MCMC approach with 500 independent chains and created posterior distributions using 5000 iterations of converged chains with a thin factor of 10. This generated a posterior distribution made with 250,000 independent samples for each parameter. The fitted parameters extracted from such posteriors can be found in Table~\ref{tab:parstarget}. 
We note that the model and data only weakly constrain the orbital period of \target\ c, $P_c$. Furthermore, posterior distributions for the semi-amplitudes of both planets, $K_b$ and $K_c$, are truncated at zero. These posteriors and their correlations are shown in Figure~\ref{fig:correlations}. 

\begin{figure}
    \centering
    \includegraphics[width=0.45\textwidth]{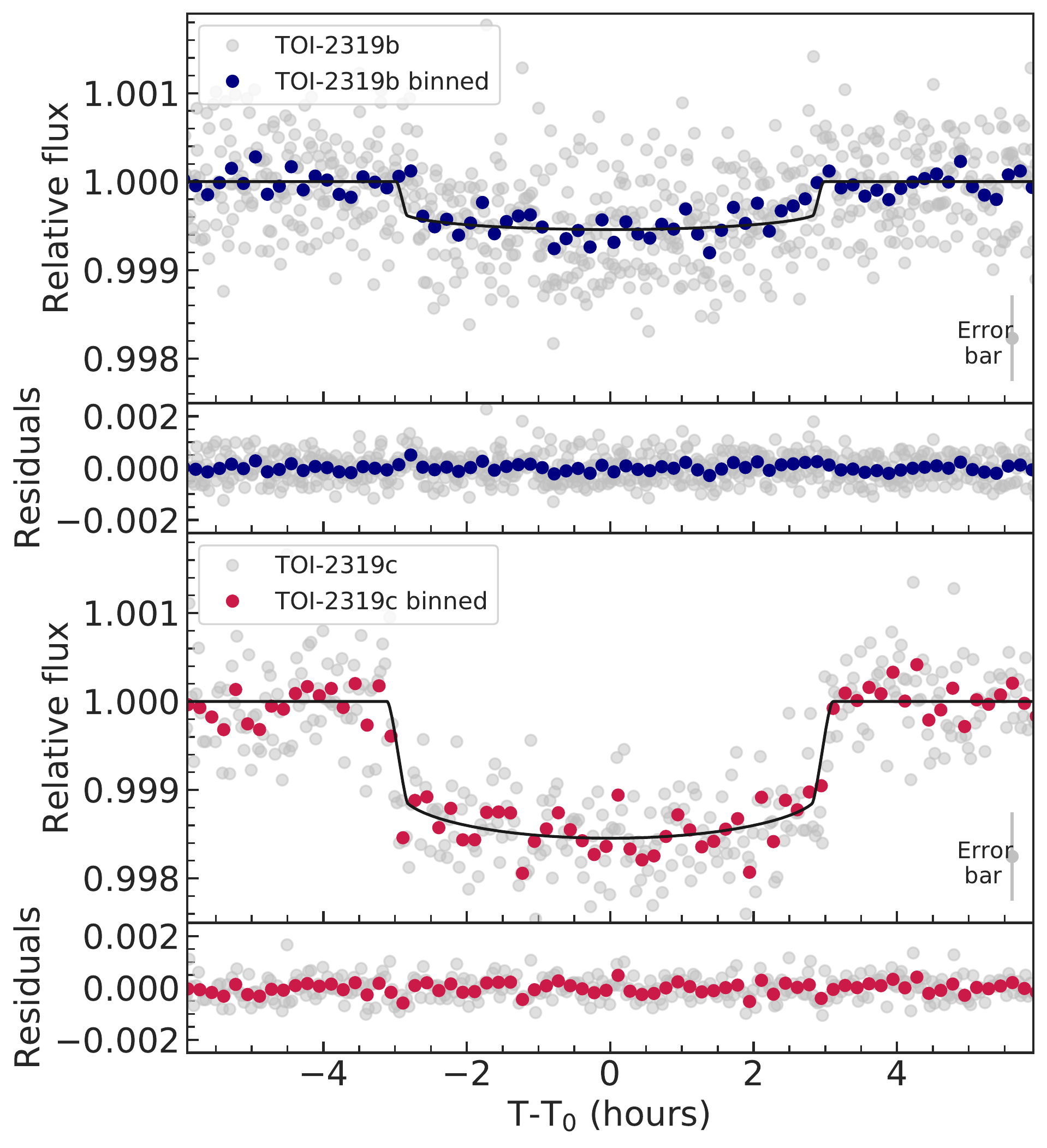}
    \caption{Phase-folded \textit{TESS} light curve of \target\,b (upper panel) and \target\,c (lower panel). 
    Nominal \tess\ data are shown in light gray together with 10-min binned data in solid colour.
    The inferred transit model for each planet is over-plotted with a solid black line. An example of the nominal white noise in the data is also shown.}
    \label{fig:transits}
\end{figure}
\begin{figure*}
    \centering
    \includegraphics[width=0.96\textwidth]{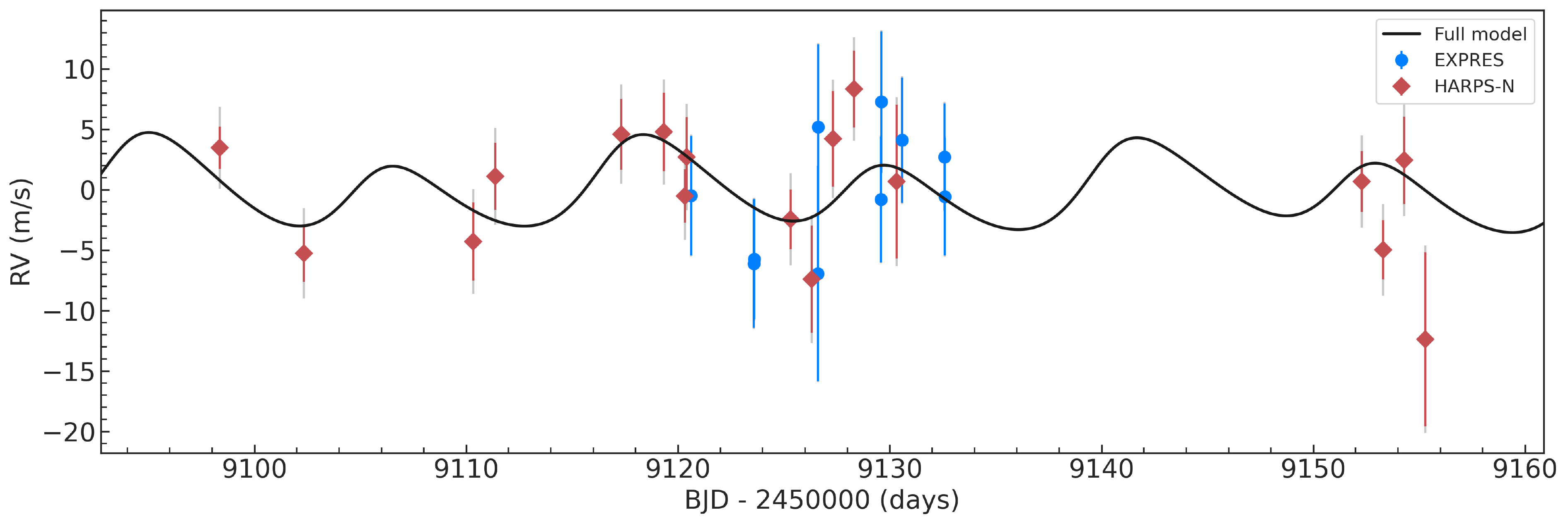} \\
    \includegraphics[width=0.48\textwidth]{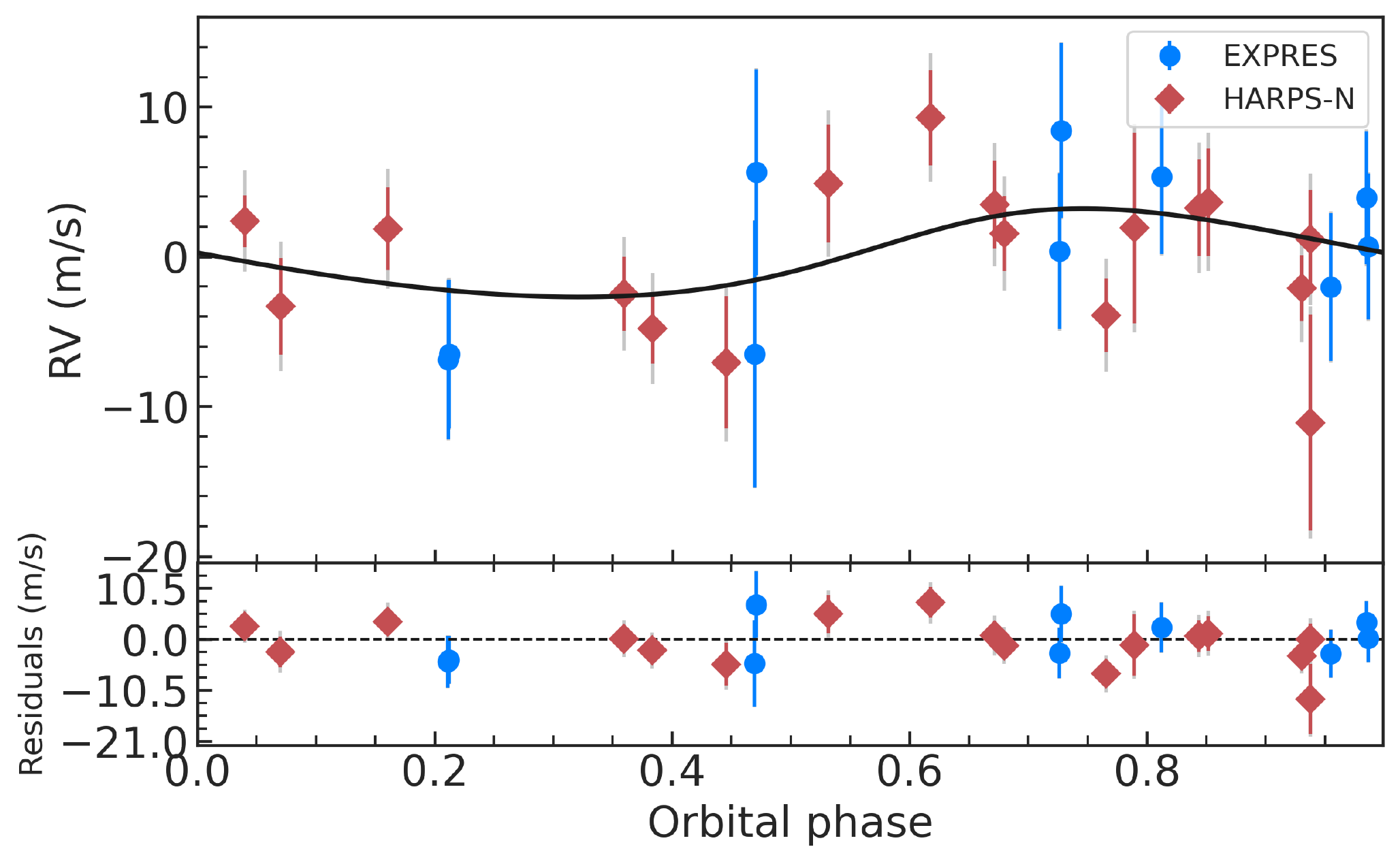}
    \includegraphics[width=0.48\textwidth]{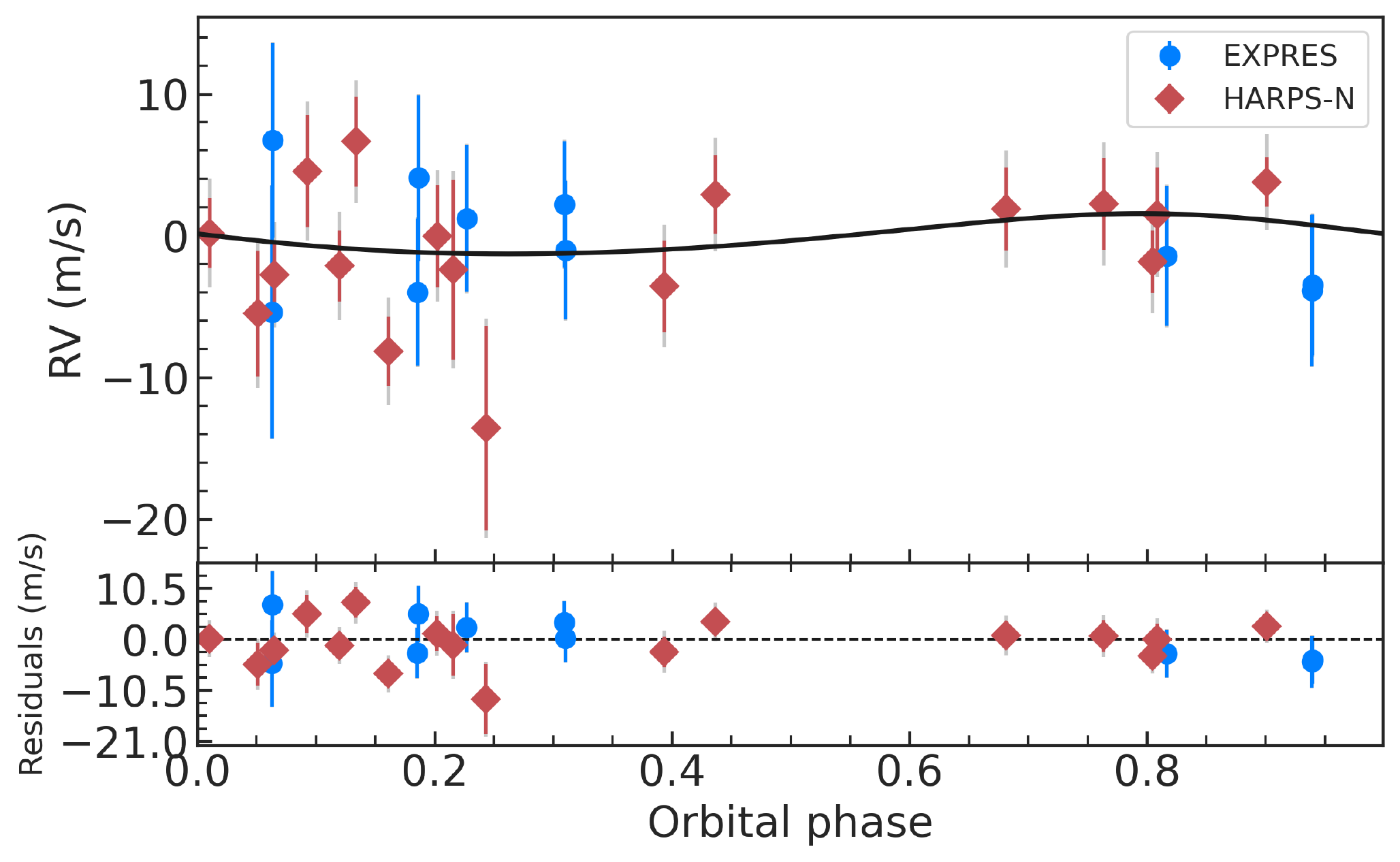}
    \caption{RV time-series (upper panel) and phase-folded RV plots for \target\, b (lower left panel) and \target\,c (lower right panel) following the subtraction of the instrumental offsets. 
    \target\,c plot has been phase folded using a period of 24.5 days.
    HARPS-N (red diamonds) and EXPRES (blue circles) RV measurements along with their nominal uncertainties are shown in each panel. The vertical grey lines mark the error bars including jitter. Solid black lines show the respective inferred model.}
    \label{fig:rvs}
\end{figure*}

The posterior of $K_b$ corresponds to a 2$\sigma$ detection, \kb, while planet~c is not detected with an upper limit of 5.6\,\ms, at 99 per cent confidence level.
Figures \ref{fig:transits} and \ref{fig:rvs} show the 
derived transit and RV models, respectively, together with the corresponding data.

\renewcommand{\arraystretch}{1.3}
\begin{table*}
\centering
  \caption{System parameters. \label{tab:parstarget}}  
  \begin{tabular}{lccl}
  \hline
  Parameter & Prior$^{(\mathrm{a})}$ & Value$^{(\mathrm{b})}$ & Comments \\
  \hline
  \multicolumn{4}{l}{\textit{Model Parameters for \target b}} \\
  \noalign{\smallskip}
    Orbital period $P_{\mathrm{orb}}$ (days)  &  $\mathcal{U}[11.5,11.7]$ & \Pb[] \\
    Transit epoch $T_0$ (BJD - 2457000)  & $\mathcal{U}[1994.25 , 1994.30]$ & \Tzerob[]  \\
    Parametrization $e \sin \omega$  &  $\mathcal{U}[-1,1]$ & \esinb & The code ensures $e < 1$   \\
    Parametrization $e \cos \omega$   &  $\mathcal{U}[-1,1]$ &\ecosb & The code ensures $e < 1$  \\
    Scaled planet radius  $R_\mathrm{p}/R_{\star}$ &  $\mathcal{U}[0,0.1]$ & \rrb[]  \\
    Impact parameter, $b$ &  $\mathcal{U}[0,1.1]$  & \bb[] \\
    Doppler semi-amplitude, $K$ (\ms) & $\mathcal{U}[0,50]$ & \kb[] & 2$\sigma$ detection \\
    \noalign{\smallskip}
    \multicolumn{4}{l}{\textit{Model Parameters for \target c}} \\
    \noalign{\smallskip}
    Orbital period $P_{\mathrm{orb}}$ (days)  &  $\mathcal{U}[  19.26 , 35]$ & \Pc[] & Truncated posterior (see Fig. \ref{fig:correlations}) \\
    Transit epoch $T_0$ (BJD - 2457000)  & $\mathcal{U}[ 2002.73 , 2002.8]$ & \Tzeroc[]  \\
    Parametrization $e \sin \omega$  &  $\mathcal{U}[-1,1]$ &  \esinc & The code ensures $e < 1$   \\
    Parametrization $e \cos \omega$   &  $\mathcal{U}[-1,1]$ & \ecosc & The code ensures $e < 1$  \\
    Scaled planet radius  $R_\mathrm{p}/R_{\star}$ &  $\mathcal{U}[0,0.1]$ & \rrc[]  \\
    Impact parameter, $b$ &  $\mathcal{U}[0,1.1]$  & \bc[] \\
    Doppler semi-amplitude, $K$ (\ms) & $\mathcal{U}[0,50]$ & 7.1 &  Upper limit (99 per cent interval of the posterior)  \\
    \multicolumn{4}{l}{\textit{Other Parameters}} \\
    \noalign{\smallskip}
    Stellar density $\rho_\star$ (${\rm g\,cm^{-3}}$) &  $\mathcal{N}[0.56,0.04]$ & \denstrb[] \\
    Parameterized limb-darkening coefficient $q_1$  & $\mathcal{U}[0,1]$ & \qone & $q_1$ parameter as in \citet{Kipping2013} \\
    Parameterized limb-darkening coefficient $q_2$ & $\mathcal{U}[0,1]$ & \qtwo & $q_2$ parameter as in \citet{Kipping2013} \\
    Offset velocity HARPS-N (\kms) & $\mathcal{U}[-0.50 , 0.50]$ & \HARPSN[] \\
    Offset velocity EXPRES (\kms) & $\mathcal{U}[-0.50 , 0.50]$ & \EXPRES[] \\
    Jitter HARPS-N (\ms) & $\mathcal{U}[0,100]$ & \jHARPSN[] \\
    Jitter EXPRES (\ms) & $\mathcal{U}[0,100]$ & \jEXPRES[] \\
    Jitter TESS (ppm) & $\mathcal{U}[0,500]$ & \jtr[] \\
    \hline
    \multicolumn{4}{l}{\textit{Derived parameters \target b}} \\
  \noalign{\smallskip}
    Planet mass ($M_{\oplus}$)  & $\cdots$ & \mpb[] & 2$\sigma$ detection \\
    Planet radius ($R_{\oplus}$)  & $\cdots$ & \rpb[] \\
    Planet density $\rho$ (${\rm g\,cm^{-3}}$) & $\cdots$ & \denpb[] \\
    Semi-major axis $a$ (AU)  & $\cdots$ & \ab[] \\
    Eccentricity $e$ & $\cdots$ & \eb[] & Upper limit of 0.72 (99 per cent interval of the posterior)  \\
    Transit duration $\tau$ (hours) & $\cdots$ & \ttotb[] \\
    Orbit inclination $i$ (deg)  & $\cdots$ & \ib[] \\
    Insolation $F_{\rm p}$ ($F_{\oplus}$)   & $\cdots$ & \insolationb[] \\
    \multicolumn{4}{l}{\textit{Derived parameters \target c}}
    \\
  \noalign{\smallskip}
    Planet mass ($M_{\oplus}$)  & $\cdots$ & \mpcup[] &  Upper limit (99 per cent interval of the posterior)  \\
    Planet radius ($R_{\oplus}$)  & $\cdots$ & \rpc[] \\
    Planet density $\rho$ (${\rm g\,cm^{-3}}$) & $\cdots$ & 0.82 & Upper limit (99 per cent interval of the posterior) \\
    Eccentricity $e$ & $\cdots$ & \ec[] & Upper limit of 0.59 (99 per cent interval of the posterior) \\
    Transit duration $\tau$ (hours) &$\cdots$ & \ttotc[] \\
    Orbit inclination $i$ (deg)  & $\cdots$ & \ic[] \\
    \hline
   \noalign{\smallskip}
  \end{tabular}
  \begin{tablenotes}\footnotesize
  \item \textit{Note} -- $^{(\mathrm{a})}$ $\mathcal{U}[a,b]$ refers to uniform priors between $a$ and $b$, $\mathcal{N}[a,b]$ to Gaussian priors with mean $a$ and standard deviation $b$.
  $^{(\mathrm{b})}$  Inferred parameters and errors are defined as the median and 68.3 per cent credible interval of the posterior distribution.
\end{tablenotes}
\end{table*}


\subsection{Statistical Validation}
The open source python package \vespa\ was used to calculate the statistical false positive probability (FPP) of both the planet candidates \citep{morton12,morton2015,Morton2016}. In brief, \vespa\ computes the probabilities of a number of astrophysical scenarios that could result in the transit events using a Bayesian framework. These consist of HEB (hierarchical eclipsing binary), EB (eclipsing binary) and BEB (background eclipsing binary). A population of stars is simulated for each scenario using the \texttt{TRILEGAL} galactic model \citep{Girardi2005} and the shape of the simulated transits compared to the transits in the observed \tess\ light curve. This results in a likelihood for each false positive scenario.

The FPPs for \planetb\ and \planetc\ are 0.05~per cent and \textless 0.001 \%, respectively, meaning that they are both below the traditionally required threshold of FPP < 1~per cent \citep{Morton2016, 2016Crossfield}. We also note that the \vespa\ model does not consider multiplicity in planet systems, which has been shown to decrease the FPP by at least an order of magnitude \citep{Lissauer2011, Lissauer2012, Lissauer2014}. \cite{Lissauer2012}, for example, estimated that systems with two or more planets in the \textit{Kepler} data were 25 times less likely to be false positives. Furthermore, the derived upper mass limits of both planets enable us to rule out that the events are caused by an eclipsing binary. As both planet candidates reach the required threshold of 99 per cent confidence level we consider both \planetb\ and \planetc\ statistically validated.


\section{Results and discussion}
\label{sec:results}

The inner planet \planetb\ (${P}_{\rm{b}}\,= $\Pb[] d) has a radius of $R_{b}$ = \rpb while the outer planet \planetc\ has a radius of $R_{c}$ = \rpc. The radial velocity measurements allowed us to constrain the mass of the innermost planet to ${M}_{{\rm{b}}}=$ \mpb\ and derive an upper mass limit of the outer planet (i.e. of planet c) of ${M}_{{\rm{c}}}<$ \mpcup. Even though the obtained spectroscopic data do not provide a $3-\sigma$ detection of the mass of either planet, the derived upper mass limits allow us to confirm that the transit signals seen in the \tess\ light curve are not the result of an eclipsing binary. Furthermore, they allow us to make predictions about future photometric and spectroscopic follow-up observations (see Sections~\ref{subsec:TTV} and \ref{subsec:RM}).


While the orbital period of the inner planet is well determined, based on the two transit events seen in the \tess\ light curve, this is not the case for the singly transiting outer planet. We, therefore, constrain ${P}_{\rm{c}}$ based on the minimum period allowed by the \tess\ light curve, the transit duration and shape, and the joint modeling of the transit and RVs. 

As shown in Figure~\ref{fig:correlations}, the joint modeling of the light curve and the RVs produce a truncated posterior distribution for ${P}_{\rm{c}}$. This distribution favours orbital periods of around 23 days. While this could indicate a 2:1 mean motion resonance (MMR) with planet b, this could also be an artefact introduced into the modeling by planet~b. Furthermore, while we can rule out orbital periods shorter than 19.26 d, it is possible that \planetc\ has an orbital period of, or close to, 19.375 d, which would be a 5:3 MMR with \planetb. The dynamical stability of these orbits and the effects of resonances in multi-planet systems is further discussion in Sections~\ref{subsec:dynamics} and ~\ref{subsec:TTV}, respectively.



\begin{figure}
    \centering
    \includegraphics[width=0.45\textwidth]{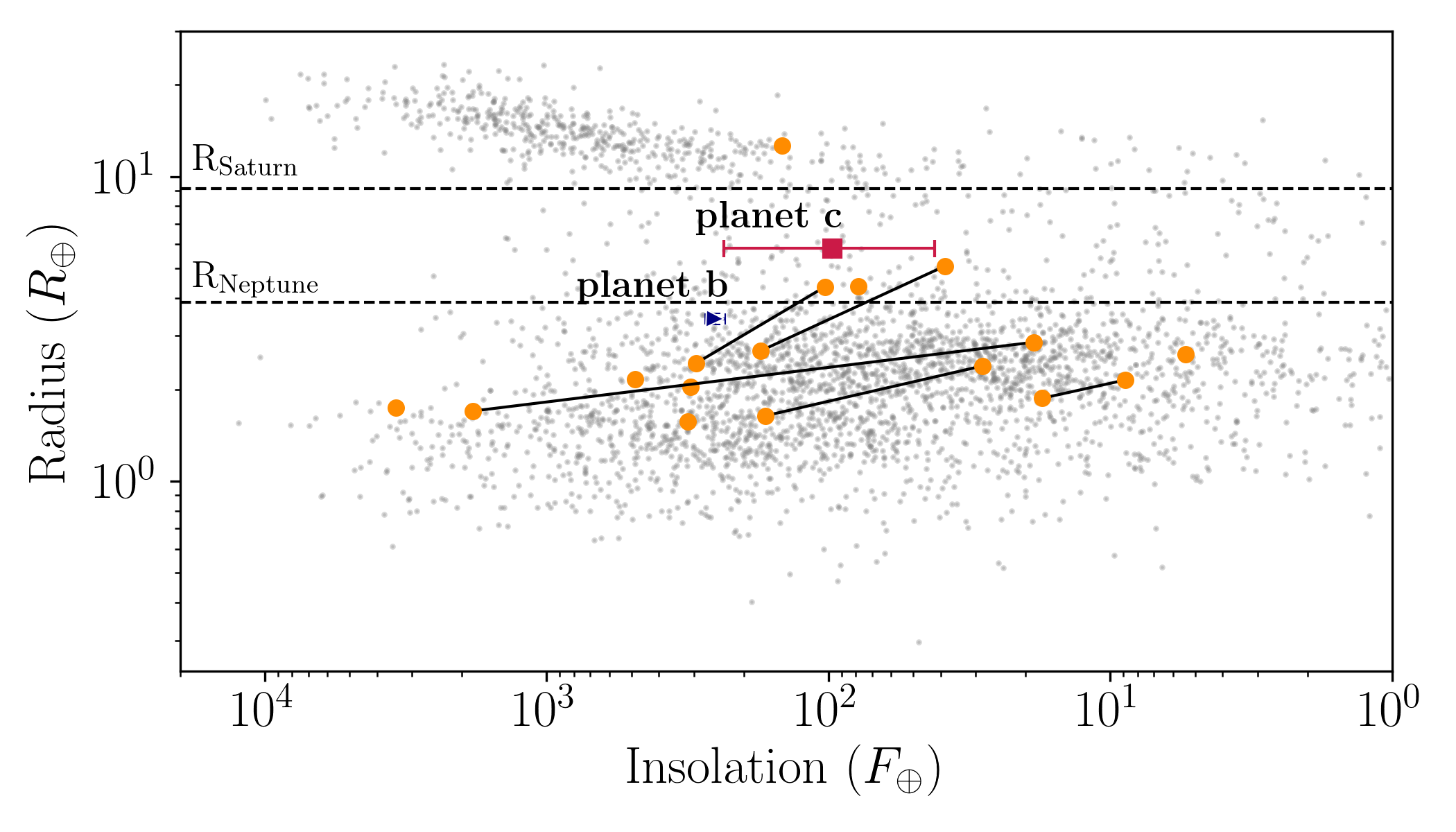}
    \caption{Planet insolation-radius diagram of confirmed exoplanets from the NASA Exoplanet Archive (grey points, retrieved \textcolor{black}{April} 2021). Orange points show members of systems with more than one planet, with mass measurements better than 50 per cent and around stars brighter than V = 10 \citep[][\textcolor{black}{see Appendix~\ref{appendixA}}]{Akeson2013}. The black lines connect planets that are within the same system. Planets that are not connected by a black line are in multi-systems where only one planet has a mass measurement with better than 50 per cent accuracy. \planetb\ and \planetc\ are shown by the blue triangle and pink square, respectively.} 
    \label{fig:radius_insolation}
\end{figure}

In order to place \target\ into a wider context, Figure~\ref{fig:radius_insolation} shows the position of planet\,b and c in the radius-insolation diagram alongside all known exoplanets (grey points). Multi-planet systems with measured masses around stars brighter than V = 10 are shown by the orange circles (\textcolor{black}{see Appendix~\ref{appendixA} for more detail on these systems}). \planetb\ and \planetc\ are depicted by the blue triangle and pink square, respectively. The figure highlights a noticeable lack of well characterised multi-planet systems around bright stars, which are key for comparative atmospheric studies. Furthermore, it shows that the planet\,c lies in a sparsely populated region of parameter space. This makes it valuable, as the characterisation of planets in this underpopulated region of parameters can help constrain theories of planet formation and evolution. 



The two planets also stand out in terms of their bulk densities. Given the minimum radius and upper mass limit of \planetc, this planet has a density < 0.82 ${\rm g\,cm^{-3}}$, suggesting that the planet has an extended gaseous envelope. Similarly, the density of \planetb\ is \denpb, making both planets prime candidates for atmospheric characterisation, as discussed further in Section~\ref{subsec:followup}. 

One possible explanation for the expected low density of planet\,c is that it formed at a greater distance from the host star prior to migrating to its current orbit. This would have allowed the planet to accrete a significant H/He envelope, due to the colder and less dense gas present farther away from the host star. Furthermore, planets that undergo this type of migration are often found to be the outer planets in MMR chains \citep{2016LeeChiang}. Future spectroscopic and photometric observations will allow us to further constrain the orbital period of planet\,c in order to determine whether the two planets are in resonance with one another.

Alternatively, the two planets could have formed \textit{in situ} and their differing planet properties resulted from subsequent diverging evolutionary pathways. For example, extreme ultraviolet irradiation from the host star could have enabled atmospheric loss through photoevaporation of the inner planet \citep[][]{Owen2016, Chen2016}, stripping it of its extended gaseous envelope, while the outer planet could have been inflated, resulting in the observed low density of planet\,c.

Theory also suggests that the low density of the planets could be due to tidal heating, which could result in an increase in entropy \cite[e.g.,][]{2019Millholland} and thus an inflated radius. Finally, \cite{2020GaoZhang} and \cite{Wang2019} independently suggest that the apparent radii could be enhanced by photochemical hazes in the atmospheres, resulting in an underestimate of the densities of planets. Future transmission spectra of planet\,c, for example at mid-infrared wavelengths where the atmosphere is less affected by hazes, will allow us to differentiate between different formation scenarios and therefore provide useful constraints for theoretical models of planet formation and migration.



\subsection{Transit Timing Variations prospects} \label{subsec:TTV}
Transit Timing Variations or TTVs are often observable in multi-planet systems as two planets dynamically interact, as predicted by \cite{Agol2005} and \cite{Holman2005}. This is especially the case when planets are near orbital resonance, which is potentially true for \target. Measuring TTVs, especially when combined with RV data allows for the refinement of planetary mass and orbital parameters, critical for interpreting atmospheric transmission spectra in smaller planets \citep[][]{Batalha2019}. It can also enable the detection of inclined non-transiting planets and can therefore lend insight into system demographics and architectures \citep[][]{Brakensiek2016}. 

TTVs were assessed for this system using the best-fit planetary parameters across a range of mass, period, and eccentricity solutions using the TTVFast framework of \textit{n}-body simulations \citep[][]{Deck2014}. Maximum likelihood solutions for the periods of planets b and c indicate a possible 2:1 resonance, which would result in TTVs with an amplitude ranging from 5-40 minutes, and a super period of approximately 2-3 years, allowing for follow-up observations to detect discernible TTVs on the scale of about a year. This amplitude would be greatly increased for non-zero eccentricities. In the window of possible period solutions, further resonant solutions include a 5:3 resonance; however, significant TTVs would not be observed away from resonance. Followup studies of this system should enable us to significantly constrain planetary masses, eccentricities, and other orbital parameters, given both the presence or absence of significant TTVs.

\subsection{Rossiter-McLaughlin effect prospects}
\label{subsec:RM}

The moderate projected rotational velocity of \target\ ($v \sin i$~$\sim$~8.2~\kms) makes it a good candidate for studying the Rossiter-McLaughlin effect \citep[RM; ][]{Rossiter24, McLaughlin24}, which provides an estimate of the spin-orbit alignment of the orbiting planets with the host star \citep[e.g.,][]{Schneider2000}. The RM effect helps to shed light onto the dynamical history of the system, as mechanisms such as planet-disk interactions help to preserve the initial spin-orbit alignment, while planet-planet interactions promote misalignment \citep[e.g.,][]{2008Chatterjee, Deeg2009, Storch2017}. The number of multi-planet systems with measured obliquities remains small \citep[e.g.,][]{2021Hjorth, 2019Dalal}. We estimate the RM effect to be \RMbLC\ and \RMcLC\ for \planetb\ and c, respectively \citep[][]{Winn2010}. Future precision RV observations (for example we obtained a typical precision of 4 \ms for this target with HARPS-N) will be able to detect the RM of planet c, thus allowing for the determination of the true obliquity of the target.


\subsection{Orbital dynamics}
\label{subsec:dynamics}
Given the uncertainty around the period of planet c, we are unable to perform a full dynamical analysis of the system, as in the work of e.g. \cite{Horner2019}. However, we can estimate the system stability by comparing the possible period scenarios of planet c to the general cases presented by \cite{Agnew2019}. 

In general, those authors found that dynamical stability can be broken into three broad regimes: highly stable orbits (when the two orbits do not approach more closely than several mutual Hill radii; and when the two orbits are more widely spaced than the 1:2 MMR); qualified stability (when the orbits are closer together than the 1:2 resonance, but have stability ensured by mutual MMR) and likely strong instability (which typically occurs for orbits that either cross, or are located closer than the 1:2 resonance, whilst not benefiting from the protection of another MMR). In this light, we consider it likely that the 23 day period estimate for planet c, and any period solution longer than that, is almost certainly a feasible, stable solution - it places that planet beyond the location of the 1:2 MMR, and so is stable so long as its eccentricity is less than $\sim0.3$ (greater than this would bring the periastron distance of planet c too close to planet b).

The minimum possible period of 19.25 days lies interior to the 1:2 resonance, and is close to the 3:5 resonance (period of 19.35 days). As can be seen in the fourth row of Figure 4 in \cite{Agnew2019}, this region is still likely to be stable, so long as the orbital eccentricity for planet c is below $\sim 0.2$.

\subsection{Feasibility of atmospheric characterisation}
\label{subsec:followup}

Known transiting multi-planetary systems with measured masses, around stars bright enough for atmospheric follow-up i.e. brighter than V = 10, are exceedingly rare. The brightness of \target\ (V = 8.855), combined with the large radii of the planets, as shown in Figure~\ref{fig:radius_insolation}, make them key targets for atmospheric characterisation via transmission spectroscopy. We assess the feasibility of such an observation using the transmission spectroscopy metric \citep[TSM; ][]{kempton18}, which provides the estimated SNR of a 10 hour observation with JWST/NIRISS \citep{doyon2012jwst}, if a cloud-free atmosphere is assumed. Based on planetary masses of 11.58 and 27.5\,$M_{\oplus}$ (Table~\ref{tab:parstarget}), and assuming a mean molecular weight of 2.3, we find the TSM to be \textcolor{black}{65 and 103}, for \planetb\ and \planetc, respectively. \textcolor{black}{The latter compares well with several of the targets currently included in JWST ERS and GTO programs, and is better than the cut-off thresholds for follow-up observations, of 96, as suggested by \cite{kempton18}. The TSM of 103 places planet\,c at least amongst the top 50 per cent of candidates suitable for atmospheric characterisation as outlined by \cite{kempton18}. Furthermore, as the planet mass used to determine this value is an upper mass limit, the TSM of planet\,c is likely to be significantly higher, likely placing it amongst the top 25 per cent of candidates best suited for atmospheric characterisation.}

\subsection{Atmospheric modelling}
\begin{figure*}
    \centering
    \includegraphics[width=0.45\textwidth]{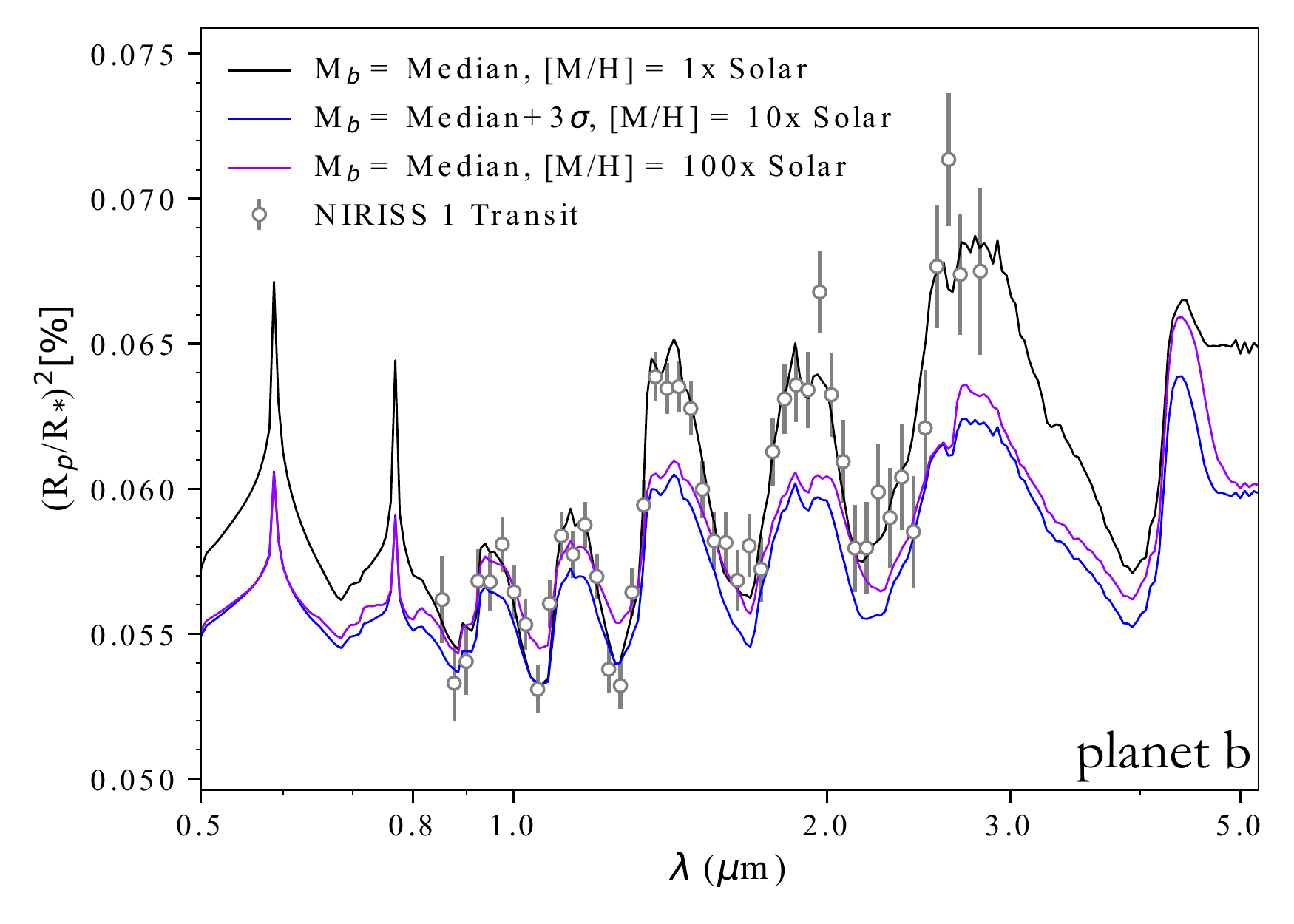}
    \includegraphics[width=0.45\textwidth]{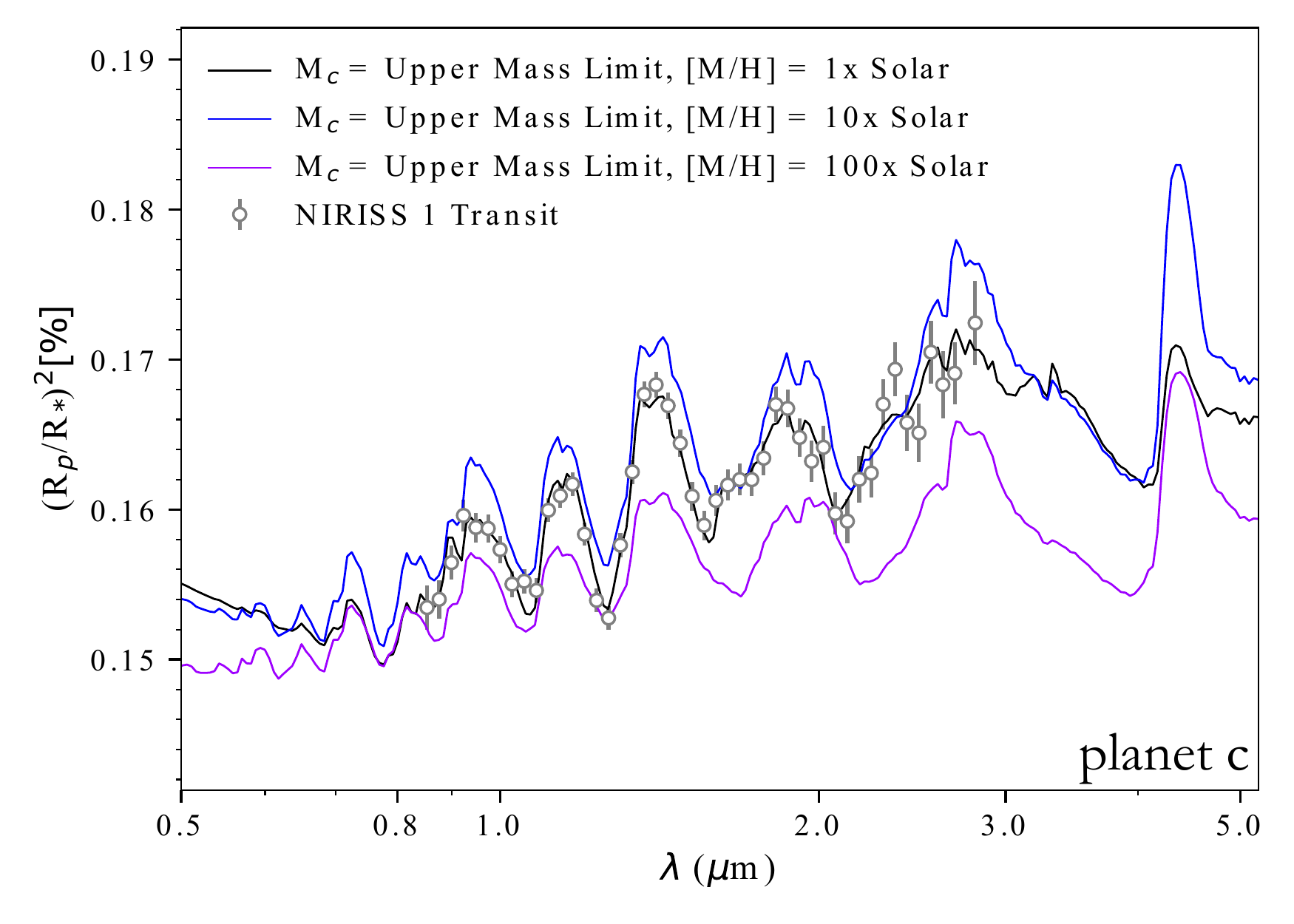}
    \caption{Models generated for planets b and c in the left and right panels, respectively. Each panel shows three models describing plausible atmospheric scenarios. In black we present an atmospheric model which has a metallicity of 1$\times$ solar, considering the RV extracted median mass and upper mass limit for planets b and c, respectively. In purple we present an atmospheric model which has a metallicity of 100$\times$ solar considering the RV extracted median mass and upper mass limit for planets b and c, respectively. In blue we present an atmospheric model which has a metallicity of 10$\times$ solar, however we consider the RV extracted median mass plus the 3$\sigma$ upper uncertainty for planet b and the upper mass limit for planet c. We overplot the simulated JWST NIRISS/SOSS observations for the 1$\times$ solar case to emphasise the precision we would obtain from a single transit observation.}
    \label{fig:atmosphere_modeling}
\end{figure*}

To assess the possibility of differentiating between different atmospheric scenarios we generated an array of forward models using the open source code {\sc chimera} \citep{Line2013} and compared these to synthetic observations of each planet which were generated using {\sc  PandExo} \citep{Batalha2017} for 1 transit observation using {\color{black}JWST} NIRISS/SOSS. A subset of these models can be seen in Figure~\ref{fig:atmosphere_modeling}. For each planet we modelled a cloud free atmosphere with an isothermal temperature profile set to the derived temperature from Table~\ref{tab:star}. For planet c we modelled the upper mass limit of 27.5 $M_{\oplus}$\ and for planet b we considered three mass scenarios: 1) the median mass, 2) the median mass + the 3$\sigma$ uncertainty and 3) the median mass - the 3$\sigma$ uncertainty. We did this so that we could capture the full range of possible transmission spectra. We then modelled the atmospheres to have a solar C/O ratio and metalicities of 1$\times$, 10$\times$ and 100$\times$ solar respectively. We used the chemical grid developed by \citet{Kreidberg2015}. In Figure~\ref{fig:atmosphere_modeling} we highlight a subset of the models. We do not show the models for scenario 3 because the lower masses would have larger observable features than the median and hence would be easier to observe. For each planet we present three models: in black we show the model for the mass and 1$\times$ solar metallicity, in purple we show the model for the mass and 100$\times$ solar and finally in blue we show the model for the mass + 3$\sigma$ and 10$\times$ metallicity. We use the mean mass and upper mass limits for planets b and c, respectively. We then overplot the predictive observations obtained from JWST NIRISS/SOSS generated using the 1$\times$ solar median mass models. The left panel, corresponding to planet\,b, shows that while with a single transit it is possible to detect the atmosphere, there remains a degeneracy between the metallicity and the mass of the planet. Future RV follow-up observations will enable us to break this degeneracy. The right panel, corresponding to planet\,c, shows that the simulated data have extremely small error bars, due to the bright star and long transit duration. These small error bars allow us to break the degeneracy between planetary mass and atmospheric metallicity. These simulations emphasise how promising these targets are for follow-up measurements and atmospheric characterisation. 

\section{Summary and conclusions}
\label{sec:conclusions}

We present the discovery of a multi-planet system (\target, TIC~349488688, TOI~2319) with a Neptune and a sub-Saturn sized planet, observed in Sector 25 of the nominal \tess\ mission. The \tess\ light curve yields two transit events for the inner planet ($P_{\color{black}b}$~=~\Pb) and a single transit event for the outer planet ($P_{\color{black}c}$~=~19.26-35 \textcolor{black}{days}). All three transit events were identified by volunteers taking part in the PHT citizen science project \citep{eisner2020method}, and the events vetted for instrumental and astrophysical false positives using the {\sc latte} vetting suite \citep{LATTE2020}. Furthermore, we statistically validated both planets using the open source software \vespa~\citep{morton12,morton2015,Morton2016} by taking into consideration the decrease in false positive probability given the multiplicity of system \citep{Lissauer2011, Lissauer2012, Lissauer2014}. 

Additionally, we obtained ground-based spectroscopic follow-up observations with HARPS-N and EXPRES in order to both constrain the orbit and planet parameters as well as to refine the stellar properties. Joint modelling of the light curve and RVs allowed us to constrain the mass of the inner planet to ${M}_{{\rm{b}}}=$ \mpb ($2-\sigma$ detection) and obtain an upper mass limit for the outer planet of ${M}_{{\rm{c}}}<$ 27.5\,$M_{\oplus}$. Furthermore, we constrained the orbit of the outer, singly-transiting planet, to be between 19.26 and 35, with the truncated model posteriors slightly favouring a period of around 23 days. This \textcolor{black}{suggests} the possibility of a 2:1 resonance with the innermost planet.

Following this, we discuss the implications of a resonance between the two planets in terms of the TTVs and show that a 2:1 resonance would result in TTVs with an amplitude between 5 and 40 minutes. We also show that the planets are suitable targets for measuring the spin-orbit alignment of the system via the RM effect, with expected amplitudes of \RMbLC\ and \RMcLC\ for \planetb\ and c, respectively. 

We also show that the properties of \planetc, which likely has an extended H/He atmosphere, combined with the brightness of the host star make it a promising targets for atmospheric characterisation. We use the TSM \citep{kempton18} to show that with a 10 hour observation with JWST/NIRISS we would obtain a SNR of 103. As an upper mass limit was used in this calculation, the value is likely to be significantly higher, making it a prime target for future atmospheric characterisation.

Finally, we generate forward models of different atmospheric compositions and compare these to synthetic observations for each planet in order to differentiate between different atmospheric scenarios. With this we show that with a single JWST NIRISS/SOSS we would be able to detect the atmospheres of these planets. Furthermore, the brightness of the star combined with the transit duration \textcolor{black}{of planet c} results in small uncertainties in the simulated spectra, which allow us to break the degeneracy between planetary mass and atmospheric metallicity \textcolor{black}{for the outer planet. Future RV follow-up observations will allow us to also break this degeneracy for planet~b.}


Overall we show that this is a very promising target for future ground and space-based follow-up observations. Continued future efforts with HARPS-N and EXPRES will be able to conclusively determine the masses of both planets and the orbital period of planet\,c, as well as search for the RM effect. Additionally, ground-based photometers, such as LCO/Sinistro \citep{LCO2013}, will allow us to observe future transit events and constrain possible TTVs, as will the space based missions such as CHEOPS \citep{broeg2013cheops}, or the upcoming PLATO mission \citep{rauer2014plato}. \target\ is also scheduled to be re-observed by the \tess\ mission during Sector 52 (May-June 2022). Finally, observations with JWST or ARIEL \citep{Tinetti2016ariel} will help to characterise the atmospheres of these scientifically valuable planets.


\section*{Data Availability}

The \tess\ data used within this article are hosted and made publicly available by the Mikulski Archive for Space Telescopes (MAST, \url{http://archive.stsci.edu/tess/}). Similarly, the Planet Hunters TESS classifications made by the citizen scientists can be found on the Planet Hunters Analysis Database (PHAD, \url{https://mast.stsci.edu/phad/}), which is also hosted by MAST. The two planet candidates and their properties have been uploaded to the Exoplanet Follow-up Observing Program for TESS (ExoFOP-TESS) website as community TOIs (cTOIs; \url{ https://exofop.ipac.caltech.edu/tess/target.php?id=349488688}).

The models of the transit events and the data validation report used for the vetting of the target were both generated using publicly available open software codes, \texttt{pyaneti} and {\sc latte}.

\section*{Acknowledgements}

We thank all of the volunteers who participated in the Planet Hunters TESS project, as without them this work would not have been possible. We also thank the editor and the referee for their comments, which improved and clarified the manuscript. Furthermore, we are very grateful to the Director of the TNG for allocating time for the HARPS-N observations from the directors discretionary time through the program ID A41DDT4.

NE also thanks the LSSTC Data Science Fellowship Program, which is funded by LSSTC, NSF Cybertraining Grant number 1829740, the Brinson Foundation, and the Moore Foundation; her participation in the program has benefited this work. AM acknowledges support from the senior Kavli Institute Fellowships. JT is a Penrose Graduate Scholar and would like to thank the Oxford Physics Endowment for Graduates (OXPEG) for funding this research. Furthermore, NE, NZ, BN and SA acknowledge support from the UK Science and Technology Facilities Council (STFC)under grant codes ST/R505006/1, ST/S505638/1 and consolidated grant no. ST/S000488. This work also received funding from the European Research Council (ERC) under the European Union’s Horizon 2020 research and innovation program (Grant agreement No. 865624).

This paper includes data collected by the \tess\ spacecraft,and we are grateful to the entire \tess\ team in obtaining and providing the lightcurves used in this analysis. Funding for the \tess\ mission is provided by the NASA Science Mission directorate. We obtained the publicly released \tess\ data from the Mikulski Archive for Space Telescopes (MAST). Resources supporting this work were also provided by the NASA High-End Computing (HEC) Program through the NASA Advanced Supercomputing (NAS) Division at Ames Research Center for the production of the SPOC data products. Furthermore, these results also made use of the Lowell Discovery Telescope at Lowell Observatory. Lowell is a private, non-profit institution dedicated to astrophysical research and public appreciation of astronomy and operates the LDT in partnership with Boston University, the University of Maryland, the University of Toledo, Northern Arizona University and Yale University. This work used the EXtreme PREcision Spectrograph (EXPRES) that was designed and commissioned at Yale with financial support by the U.S. National Science Foundation under MRI-1429365 and ATI-1509436 (PI D. Fischer). Finally, the research leading to these results has partially received funding from the KU~Leuven Research Council (grant C16/18/005: PARADISE), from the Research Foundation Flanders (FWO) under grant agreement G0H5416N (ERC Runner Up Project), as well as from the BELgian federal Science Policy Office (BELSPO) through PRODEX grant PLATO.

Finally, NE and OB wish to thank the Asterix comics, which provided the inspiration for our in-house nickname for this planet system of \textit{Idefix}.

This research made use of Astropy, a community-developed core Python package for Astronomy \citep{astropy2013}, matplotlib \citep{matplotlib}, pandas \citep{pandas}, NumPy \citep{numpy}, astroquery \citep{ginsburg2019astroquery} and sklearn \citep{pedregosa2011scikit}.




\bibliographystyle{mnras}
\bibliography{bibs} 



\appendix
\section{Confirmed multi-planet systems}
\label{appendixA}

\renewcommand{\arraystretch}{1.3}
\begin{table*}
\centering
  \caption{Bright multi-planet system. \label{tab:multistars}}  
  \begin{tabular}{llllllll}
  \hline
    Host name	&	Planet letter	&	\textcolor{black}{R$_{\mathrm{pl}}$} (R$_{\oplus}$)	&	\textcolor{black}{M$_{\mathrm{pl}}$} (M$_{\oplus}$)	&	P$_\mathrm{{pl}}$ (days)			&	Vmag		& No. confirmed planets	&	Reference \\
    \hline
    GJ 143		&	b				&	$2.61 _{-0.16}^ {0.17}$				&	$22.7 _{-1.9}^ {2.2}$				&	$35.61253 _{-0.00062}^ {0.0006}$	&	8.08		& 2						&	\cite{2019Dragomir} \\
    HAT-P-11	&	b				&	$4.36 _{-0.06}^ {0.06}$				&	$26.698 _{-2.22}^ {2.22}$			&	$4.8878$							&	9.46		& 2						&	\cite{2018Yee} \\
    HD 106315	&	b				&	$2.44 _{-0.17}^ {0.17}$				&	$12.6 _{-3.2}^ {3.2}$				&	$9.55237 _{-0.00089}^ {0.00089}$	&	8.951		& 2						&	\cite{2017Barros} \\
            	&	c				&	$4.35 _{-0.23}^ {0.23}$				&	$15.2 _{-3.7}^ {3.7}$				&	$21.05704 _{-0.00046}^ {0.00046}$	&	    		&  						&	                     \\
    HD 15337	&	b				&	$1.64 _{-0.06}^ {0.06}$				&	$7.51 _{-1.01}^ {1.09}$				&	$4.75615 _{-0.00017}^ {0.00017}$	&	9.1			& 2						&	\cite{2019Gandolfi} \\
        	    &	c				&	$2.39 _{-0.12}^ {0.12}$				&	$8.11 _{-1.69}^ {1.82}$				&	$17.1784 _{-0.0016}^ {0.0016}$		&   			& 						&	     \\
    HD 213885	&	b				&	$1.745 _{-0.05}^ {0.05}$			&	$8.83 _{-0.65}^ {0.66}$				&	$1.00804 _{-0.00002}^ {0.00002}$	&	7.95		& 2						&	\cite{2020Espinoza} \\
    HD 23472	&	b				&	$1.872 _{-1.32}^ {1.32}$			&	$17.92 _{-14.0}^ {1.41}$			&	$17.667 _{-0.095}^ {0.142}$			&	9.73		& 2						&	\cite{2019Trifonov} \\
        	    &	c				&	$2.149 _{-0.34}^ {0.34}$			&	$17.18 _{-13.77}^ {1.07}$			&	$29.625 _{-0.171}^ {0.224}$			&	    		& 						&	                 \\
    HD 3167		&	b				&	$1.7 _{-0.08}^ {0.08}$				&	$5.02 _{-0.38}^ {0.38}$				&	$0.95962 _{-0.00003}^ {0.00003}$	&	8.97		& 3						&	\cite{2017Christiansen} \\
        		&	c				&	$2.86 _{-0.22}^ {0.22}$				&	$9.8 _{-1.24}^ {1.3}$				&	$29.83832 _{-0.0032}^ {0.00291}$	&	    		&  						&	     \\
    HD 39091	&	c				&	$2.042 _{-0.05}^ {0.05}$			&	$4.82 _{-0.86}^ {0.84}$				&	$6.2679 _{-0.00046}^ {0.00046}$		&	5.65		& 2						&	\cite{2018Huang} \\
    HD 86226	&	c				&	$2.16 _{-0.08}^ {0.08}$				&	$7.25 _{-1.12}^ {1.19}$				&	$3.98442 _{-0.00018}^ {0.00018}$	&	7.93		& 2						&	\cite{2020Teske} \\
    Kepler-93	&	b				&	$1.569 _{-0.11}^ {0.11}$			&	$4.544 _{-0.85}^ {0.85}$			&	$4.72674 _{-0.000001}^ {0.000001}$	&	9.996		& 2						&	\cite{2015Dressing} \\
    TOI-421		&	b				&	$2.68 _{-0.18}^ {0.19}$				&	$7.17 _{-0.66}^ {0.66}$				&	$5.19672 _{-0.00049}^ {0.00049}$	&	9.931		& 2						&	\cite{Carleo2020} \\
        		&	c				&	$5.09 _{-0.15}^ {0.16}$				&	$16.42 _{-1.04}^ {1.06}$			&	$16.06819 _{-0.00035}^ {0.00035}$	&	    		& 						&	 \\
    WASP-8		&	b				&	$12.666 _{-0.56}^ {0.56}$			&	$807.288 _{-104.88}^ {104.88}$		&	$8.15872 _{-0.00001}^ {0.00001}$	&	9.789		& 2						&	\cite{2010Queloz} \\
     \hline
   \noalign{\smallskip}
  \end{tabular}
   \begin{tablenotes}\footnotesize
  \item \textit{Note} -- Confirmed exoplanets from the NASA Exoplanet Archive that are members of systems with more than one planet, with mass measurements better than 50 per cent and around stars brighter than V = 10 \citep{Akeson2013}. All parameters are as listed in the NASA Exoplanet Archive as of April 2021.
\end{tablenotes}
\end{table*}



\bsp	
\label{lastpage}
\end{document}